\documentclass[12pt]{article}
\usepackage[left=2cm, right=2cm, top=2cm, bottom=2cm]{geometry}
\usepackage{amsmath, amssymb, amsthm}
\usepackage{graphicx}
\usepackage{array}
\usepackage{abstract, sectsty, natbib}
\usepackage{setspace, bm, bbm, url, tikz}
\usepackage{algorithm, algorithmic}
\usepackage[hang, small, bf]{caption}
\usepackage{booktabs, subfigure, enumitem}
\usepackage{abstract,bigints, color}
\usepackage{algorithm, algorithmic}

\usetikzlibrary{shapes,shapes.geometric,arrows,fit,calc,positioning,automata,}

\providecommand{\keywords}[1]{\hspace{-0.5cm}Keywords: #1}
\providecommand{\summary}[1]{\textbf{Summary}: #1}
\newcommand\Tstrut{\rule{0pt}{2.5ex}}  
\newcommand\Bstrut{\rule[-1.5ex]{0pt}{0pt}}

\title{Bayesian Spatial Monotonic Multiple Regression}
\author{Christian Rohrbeck$^1$\thanks{Address for correspondence: Christian Rohrbeck, STOR-i CDT, Old Engineering Building, Lancaster University, Lancaster, LA1 4YF, UNITED KINGDOM, Email: \texttt{c.rohrbeck@lancaster.ac.uk}} , Deborah Costain$^2$, Arnoldo Frigessi$^{3}$}
\date{{\small$^1$STOR-i Centre of Doctoral Training, Lancaster University, $^2$Department of Mathematics and Statistics, Lancaster University, $^3$Oslo Centre for Biostatistics and Epidemiology}}

\begin{document}

\onehalfspacing
\maketitle

\summary{We consider monotonic, multiple regression for a set of contiguous regions (lattice data). The regression functions permissibly vary between regions and exhibit geographical structure. We develop new Bayesian non-parametric methodology which allows for both continuous and discontinuous functional shapes and which are estimated using marked point processes and reversible jump Markov Chain Monte Carlo techniques. Geographical dependency is incorporated by a flexible prior distribution; the parametrisation allows the dependency to vary with functional level. The approach is tuned using Bayesian global optimization and cross-validation. Estimates enable variable selection, threshold detection and prediction as well as the extrapolation of the regression function. Performance and flexibility of our approach is illustrated by simulation studies and an application to a Norwegian insurance data set.}

\vspace{0.3cm}


\section{Introduction}

Geospatial data are considered in several areas, including ecology \citep{Guttorp1991}, forestry \citep{Penttinen1992} and epidemiology \citep{Waller2004}. Data in a locally aggregated form, \textit{lattice data} \citep{Cressie1993}, are common due to practicality or confidentiality concerns and are typically over an irregular lattice. Statistical methods for such area-level data aim to model associations between a response variable and a set of explanatory variables, whilst accounting for potential geographical dependency in the model parameters. To introduce geographical dependence, a neighbourhood structure, often based upon the arrangement of the areal units (regions) on a map, is typically introduced.

Although widely applied, most approaches only consider geographical variation in the baseline or random effect and are otherwise usually limited to the regression function being linear \citep{Waller2010}. From this perspective, a modelling framework enabling both geographical variation and non-linearity in the regression surface would increase flexibility. In this paper, the constraint of linearity is thus relaxed and substituted by one of monotonicity, an important assumption in several applications \citep{Royston2000, Koushanfar2006, Farah2013, Wilson2014}. Indeed, tests of monotonicity for the underlying process are introduced by \citet{Bowman1998} and \citet{Ghosal2000}. Conditional on the monotonicity constraint, we develop methodology which allows for estimation of the association between the response and explanatory variables for each region; whilst exploiting any neighbourhood structure. A brief review of the motivating work, with limitations, follows.

The estimation of a multivariate, monotonic function is considered in several statistical areas and is usually referred to as \textit{isotonic regression}. Early publications discuss the inference on parameter values under monotonic constraints \citep{Ayer1955, Brunk1955, BarlowBrunk1972}. Algorithms for solving these problems are available in the optimization literature \citep{Brunk1957, Qian1992, Luss2012} with the derived solutions being of piecewise constant form. Isotonic regression is further considered for additive \citep{Bacchetti1989, Tutz2007, Cheng2009} and high-dimensional models \citep{Fang2012, Bergersen2014}, functional data analysis \citep{Ramsay2006, Bornkamp2009} and Bayesian non-parametrics \citep{Holmes2003, Shively2009, Saarela2011, Lin2014}. Despite this variety of approaches, isotonic regression is rarely applied to geospatial data. One of the few examples is the work by \citet{MortonJones2000} using additive modelling with univariate monotonic functions in an epidemiological setting. Note, that possible geographical auto-correlation between functions is not allowed for in their work.

Geographical variation of the regression function, on the other hand, is usually considered in a generalized linear, or additive, modelling framework. Geographically weighted regression (GWR) \citep{Brunsdon1998, Fotheringham2003} is the dominant approach and, for instance, applied in forestry \citep{Zhang2004} and social science \citep{Cahill2007}. Whilst GWR is based on weighted least-squares methodology, the geographically varying coefficient (GVC) model \citep{Assuncao2003} 'borrows' information locally via a Bayesian specification and conditional autoregressive (CAR) modelling \citep{Besag1974, Besag1991}. \citet{Waller2007} find that both GWR and GVC are qualitatively similar by applying them to alcohol and violence data. \citet{Scheel2013} introduce an alternative approach which borrows statistical information locally for variable selection rather than for the estimation of covariate effects. For the more flexible class of generalized additive models, \citet{Congdon2006} proposes the decomposition into local and geographically filtered effects. However, none of these methods offer flexibility in terms of recovering discontinuities. More precisely, abrupt changes in the regression surface are not captured unless these are explicitly included. Negligence of such effects may result in a bias due to oversmoothing \citep{Bowman1997}. In the rest of this paper, we refer to these discontinuities as threshold effects.

We introduce a novel Bayesian, non-parametric, methodology, \textit{Bayesian Spatial Monotonic Multiple Regression} (BSMMR), to facilitate analysis of lattice data under the sole assumption of response function monotonicity in the covariates. Extending the approach of \citet{Saarela2011}, the regional (areal) monotonic functions are each represented by a set of marked point processes. The point process formulation is highly flexible and permits both smooth contours and threshold effects in the regression surface. Beliefs on the geographical similarity of the monotonic functions are incorporated by a joint prior distribution. In particular, the prior is constructed based upon a pair-wise discrepancy measure, resulting in a Gibbs distribution on functional spaces. The defined prior is flexible in the sense that dependency between functions may be either constant, increasing or decreasing with an increasing functional mean. In order to tune the prior, we propose a new algorithm, \textit{EGO-CV}, which combines the concepts of cross-validation and Bayesian global optimization. Realizations of the posterior are obtained by a reversible jump MCMC (RJMCMC) algorithm \citep{Green1995}. Stored samples facilitate the analysis of the regression surface with regard to threshold effects, variable selection, prediction and extrapolation. 

The remainder of this paper is organized as follows: Section 2 details the statistical framework for the BSMMR approach, focusing on the point process representation of the regional monotonic regression functions and the constructed prior density. The algorithm for estimating the functions is outlined at the end of Section 2 and further details are provided in Appendix A. In order to assess the performance and sensitivity of the RJMCMC algorithm, BSMMR is then applied to simulated data in Section 3. In Section 4, our methodology is applied to Norwegian insurance and meteorological data, with a view to investigating weather related claim dynamics over the region. Finally, the paper concludes with a summary and discussion of our approach in Section 5.
\section{Modelling and Inference}
\label{sec:theory}

BSMMR is derived from several subcomponents which are explained in this section. Subsection \ref{sec:probmodel} outlines the areal data construct and also the notation used. More specifically, the modelling framework for geographical variation in the functional relationship between the response and explanatory variables is formalised for a finite set of regions. Geographical proximity of the regions is specified, for instance, by an adjacency matrix. Without loss of generality, the monotonic functions are set to be isotonic: any monotonic function can be made isotonic by reversing some of the coordinate axis. Subsection \ref{sec:pointprocess} summarises the marked point process representation for a single monotonic function (as introduced by \citet{Saarela2011}) and which is extended in this work to lattice data: allowing for geographical variation. This representation of the monotonic functions is then embedded in a Bayesian framework that facilitates borrowing statistical information between regions. Construction, with motivation, of the joint prior on the monotonic functional spaces is then detailed in Subsection \ref{sec:prior}. Posterior realisations and estimates of the model parameters are obtained by the algorithm described in Subsection \ref{sec:Inference}. 

\subsection{Probability model and Notation}
\label{sec:probmodel}

Suppose data are available in the form of a lattice (regular or irregular) for a set of $K$ contiguous regions. Let $y_k\in\mathbb{R}$ and $\mathbf{x}_k = \left(x_{k,1},\ldots, x_{k,m}\right)\in X_k\subset\mathbb{R}^m$ denote the response and explanatory variables, respectively, for region $k$, $k=1,\ldots,K$. The closed set $X_k$ is the regional covariate space which is permissibly different across regions. The associated probability model is then formally defined as
\begin{equation}
f(y_k~|~\lambda_k(\mathbf{x}_k), \bm{\theta}_k),
\label{eq:probmodel}
\end{equation}
where $\lambda_k(\cdot)$ refers to the monotonic function for region $k$. In this work, each $\lambda_k(\cdot)$ is assumed to lie within a prespecified interval $\left[\delta_{\min}, \delta_{\max}\right]$. Monotonicity is defined in terms of the partial Euclidean ordering, that is, for any two covariate values $\mathbf{u},\mathbf{v}\in X_k$ with $\mathbf{u}\leq\mathbf{v}$ component-wise, $\lambda_k(\mathbf{u})\leq \lambda_k(\mathbf{v})$. The vector $\bm{\theta}_k$ contains additional model parameters which may also vary geographically. Note, the probability model \eqref{eq:probmodel} contains generalisations of several modelling frameworks, for example, the generalised linear modelling family with the linear predictor being replaced here by $\lambda_k$.

\subsection{A marked point process model formulation}
\label{sec:pointprocess}

The formulation by \citet{Saarela2011} is applied respectively to each of the $K$ monotonic functions. Hence, $\lambda_k$, $k=1,\ldots,K$, is postulated to be piecewise constant and is represented by a marked point process, $\Delta_k$, on $X_k$. We assume $X_k = \left[0,1\right]^m$ for notational simplicity in the rest of this subsection. The points in $\Delta_k$ define the locations in $X_k$ of the changes in the functional levels of $\lambda_k$. Consequently, the estimation problem is shifted from the space of monotonic functions to one of marked point processes with monotonic constraints. The point process approach does not restrict the space of potential solutions, as any bounded function can be approximated up to a desired degree of accuracy by increasing the number of support points in $\Delta_k$. 

\begin{figure}
\begin{center}
\includegraphics[height=7.2cm]{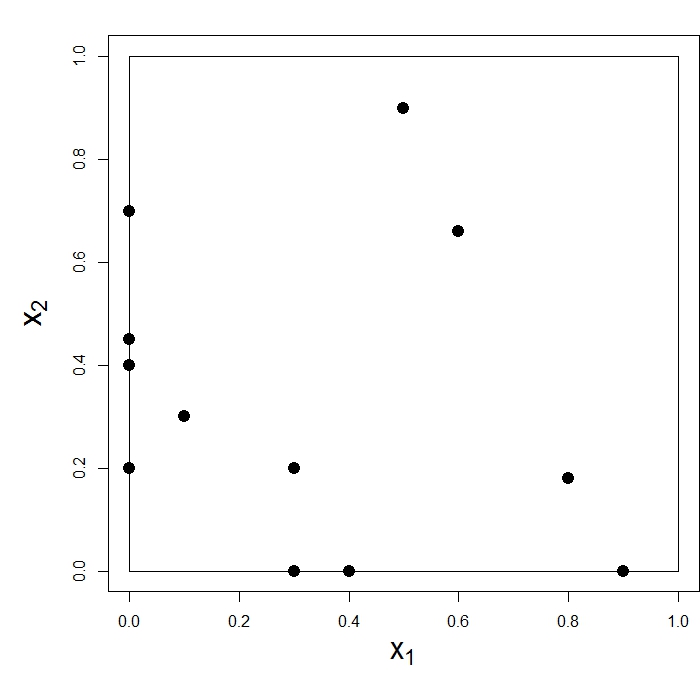}
\hspace{0.5cm}
\includegraphics[height=7.2cm]{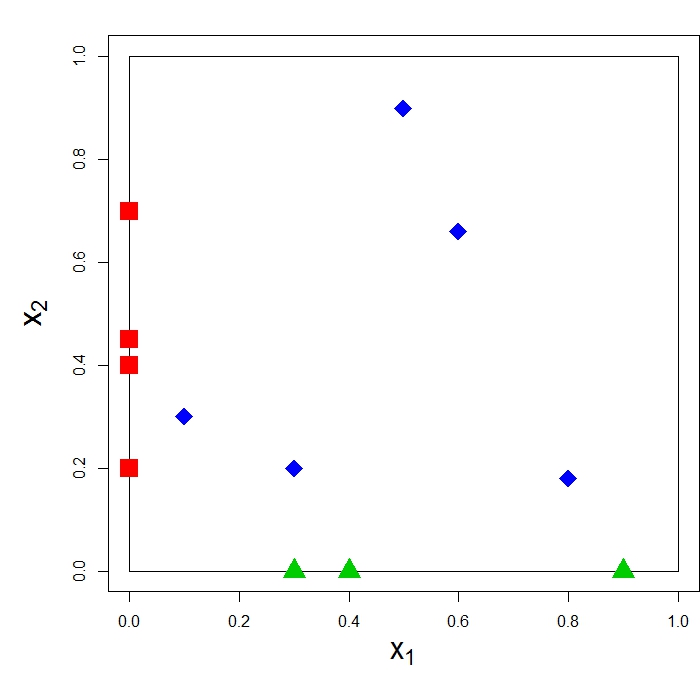}
\caption{Partition of $\Delta_k$ (left) into $\bm{\Delta}_k = \left\{\Delta_{k,1}, \Delta_{k,2} , \Delta_{k,3}\right\}$ (right) for the bivariate covariate space $[0,1]^2$. In the right panel, the marked point processes $\Delta_{k,1}$ (triangle) and $\Delta_{k,2}$ (square) are defined on the one-dimensional covariate subsets, $\left\{1\right\}$ and $\left\{2\right\}$, respectively. $\Delta_{k,3}$ (diamond) is defined on the full covariate set $\left\{1,2\right\}$.}
\label{fig:MarkedPointProcess}
\end{center}
\end{figure}

\citet{Saarela2011} further propose a partition of $\Delta_k$ into a set $\bm{\Delta}_k = \left\{\Delta_{k,i},~i=1,\ldots,I\right\}$, with the marked point processes being defined on disjoint subspaces of $X_k$. The marked point process $\Delta_{k,i}$ on subspace $i$ for region $k$ is then formally denoted by
\begin{equation}
\Delta_{k,i} = \left\{\left(\bm{\xi}_{k,i,j}, \delta_{k,i,j}\right) ~ : ~ j = 1,\ldots,n(\Delta_{k,i})\right\},
\label{eq:representation}
\end{equation}
where $\bm{\xi}_{k,i,j}$ and $\delta_{k,i,j}$ refer to a location and associated mark, respectively, and $n(\Delta_{k,i})$ is the number of points in process $\Delta_{k,i}$. In this paper, the marked point processes are defined on the non-empty subsets of the covariate set. For instance, $\Delta_{k,1}$ is based on the first covariate only and contains the locations with all but the first component being 0, $\bm{\xi}_{k,1,j} = \left(\cdot,0,\ldots,0\right)$, $j = 1,\ldots,n\left(\Delta_{k,1}\right)$. Assumed $m=2$, this partitioning results in $I=3$ subprocesses, see Figure \ref{fig:MarkedPointProcess}, two for the one-dimensional subsets and one for the full covariate set.

The functional level $\lambda_k(\mathbf{x})$ is then defined by $\bm{\Delta}_k$ as the highest mark $\delta_{k,i,j}$ such that $\mathbf{x}$ imposes a monotonic constraint on the associated location $\bm{\xi}_{k,i,j}$. Formally, $\lambda_k(\mathbf{x})$ results in 
\begin{equation}
\lambda_k(\mathbf{x}) = \max_{i,j} \left\{\delta_{k,i,j}: \bm{\xi}_{k,i,j}\preceq\mathbf{x}\right\},
\label{eq:level}
\end{equation}
where $\preceq$ denotes the partial Euclidean ordering. This leads indeed to a monotonic function as shown by \citet{Saarela2011}. The reader is referred to \citet{Saarela2011} for a discussion on other potential choices for the definition of $\lambda_k$.

The partition $\bm{\Delta}_k$ in \eqref{eq:representation} simplifies the analysis of the monotonic function. Most importantly, in the context of parsimony, investigating the estimated marked point processes allows for variable selection. Suppose that, for instance, the first explanatory variable for region $k$, $x_{k,1}$, is redundant. Consequently, the functional level $\lambda_k(\mathbf{x})$ will be constant with increasing $x_{k,1}$, i.e.\ $\lambda_k(\mathbf{x}) = \lambda_k(\mathbf{x} + \bm{\epsilon})$, $\forall ~\mathbf{x}\in X_k$, where $\bm{\epsilon} = \left(\epsilon, 0,\ldots,0\right)$ has positive first component and is zero otherwise. The points in $\Delta_k$ are hence 0 in the first component, as they represent the locations of the changes in the functional level. In the bivariate case, this redundancy implies that all locations lie on the vertical $x_2$-axis in Figure \ref{fig:MarkedPointProcess}. Therefore, all points are contained in $\Delta_{k,2}$ after the partition of $\Delta_k$ into $\bm{\Delta}_k$. More generally, for higher dimensions, subprocesses considering point locations with non-zero first component are empty conditional on the partitioning.

\subsection{Modelling geographical dependency}
\label{sec:prior}

Beliefs on geographical dependencies in $\lambda_1,\ldots,\lambda_K$ are accommodated by a joint prior distribution. Little research exists on such models for both monotonic functions and marked point processes with monotonic constraints. A prior is therefore constructed in this subsection which focuses on monotonic functions. This choice is based upon the assumption that the main interest lies in the functional shapes. More precisely, the prior model penalises discrepancies in the functional levels of $\lambda_k$ and $\lambda_{k'}$, $k\neq k'$, $k,k'\in\left\{1,\ldots, K\right\}$, and not the number and location of support points.

\begin{figure}
\begin{center}
\subfigure[]{\includegraphics[width=8cm]{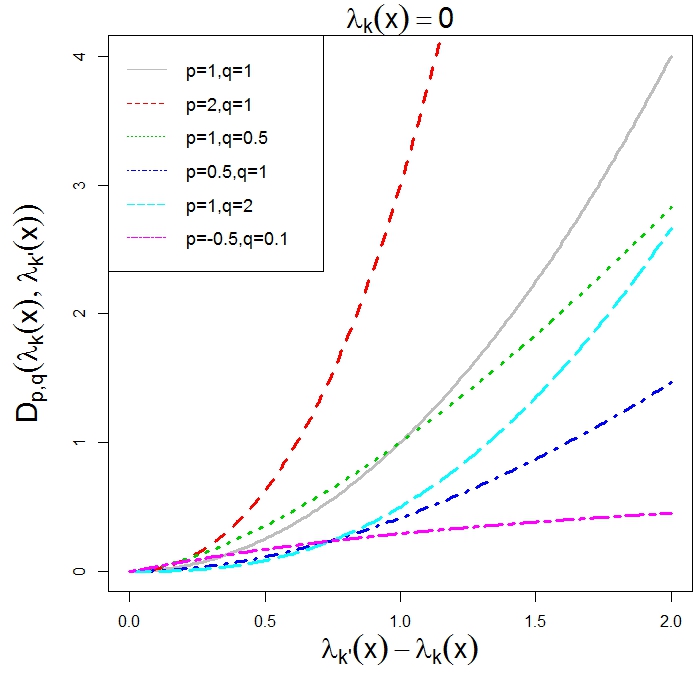}\label{fig:prior1}}
\hspace{1cm}
\subfigure[]{\includegraphics[width=8cm]{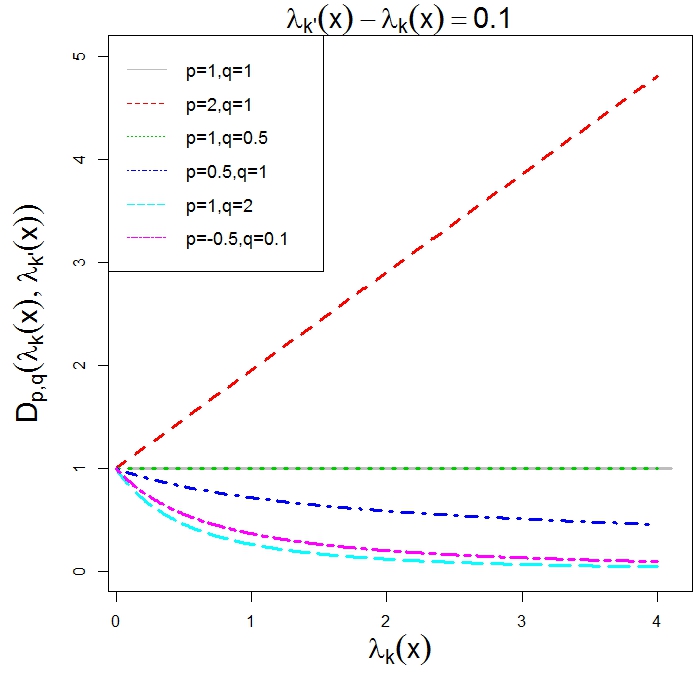}\label{fig:prior2}}
\caption{Pointwise behaviour of the discrepancy $D_{p,q}$ for different values of $p$ and $q$ with (a) increasing difference $\lambda_{k'}(\mathbf{x})- \lambda_{k}(\mathbf{x})$ in the functional means but constant functional level for $\lambda_k(\mathbf{x})$ and (b) constant difference $\lambda_{k'}(\mathbf{x})- \lambda_{k}(\mathbf{x}) = 0.1$ but increasing functional level of $\lambda_k(\mathbf{x})$.}
\label{fig:behaviiourpq}
\end{center}
\end{figure}

In the first step, a pair-wise discrepancy measure $D$ for two functions $\lambda_k$ and $\lambda_{k'}$ is introduced. For notational simplicity, the levels of both functions are assumed to be non-negative since in general, one would naturally consider $\lambda_k(\mathbf{x})-\delta_{\min}$ instead of $\lambda_k(\mathbf{x})$. Also, $D$ should be minimal, if and only if, the functions are equal. In certain applications, differences in the lower or higher functional levels should be particularly avoided. These considerations result in the discrepancy measure $D$ formally being defined by 
\begin{equation}
D_{p,q}(\lambda_{k}, \lambda_{k'}) = \int_{A_{k,k'}} \Big|\left[1+\lambda_k(\mathbf{x})\right]^p - \left[1+\lambda_{k'}(\mathbf{x})\right]^p\Big| \left|\lambda_k(\mathbf{x}) - \lambda_{k'}(\mathbf{x})\right|^q ~\mbox{d}\mathbf{x}~,~ p\in\mathbb{R}~,~q\geq0.
\label{eq:D}
\end{equation}
The integral in \eqref{eq:D} can be computed efficiently in our setting as $\lambda_k$ and $\lambda_{k'}$ are taken to be piecewise constant (Subsection \ref{sec:pointprocess}). The functional levels in the first modulus term are increased by 1 in order to ensure numerical stability for the case $p<0$ as $\lambda_k(\mathbf{x})$ can be close to 0. Figure \ref{fig:prior1} illustrates that the discrepancy increases with increasing difference $\lambda_{k'}(\mathbf{x})- \lambda_{k}(\mathbf{x})$ in the functional levels regardless of the values for $p$ and $q$. Further, Figure \ref{fig:prior2} shows that the discrepancy for a difference in the functional levels increases with increasing functional levels for $p>1$ while it decreases for $p<1$ and remains constant for $p=1$. For instance, in Figure \ref{fig:prior2}, the setting $p=2, q=1$ leads to a five-fold discrepancy increase for a level difference $\lambda_{k'}(\mathbf{x}) - \lambda_{k}(\mathbf{x})=0.1$ when $\lambda_k(\mathbf{x}) = 4$ compared to when $\lambda_k(\mathbf{x})=0$. In contrast, a value of $p<0$ leads to a reduction in the discrepancy for higher values of $\lambda_k(\mathbf{x})$. 

Sensitivity on $p$ and $q$ is explored via simulations in Section \ref{sec:SimGauss}. The domain $A_{k,k'}$ depends on the covariate spaces for regions $k$ and $k'$ and two possible settings are considered in Section \ref{sec:SimBinom}. The first defines $A_{k,k'}$ as the set of values contained in both covariate spaces, so that $A_{k,k'}$ is the intersection of $X_k$ and $X_{k'}$: $A_{k,k'}=X_k\cap X_{k'}$, while the second defines it as the union of $X_k$ and $X_{k'}$: $A_{k,k'}=X_k\cup X_{k'}$. The latter facilitates the extrapolation of $\lambda_k$ to covariate values contained in $X_{k'}$ and vice versa.

The joint prior on $\lambda_1, \ldots, \lambda_K$ is then defined as a Gibbs distribution with the discrepancy measure $D_{p,q}$ in \eqref{eq:D} as a per-potential. In order to avoid overfitting, the model is extended to accommodate model complexity via a Geometric prior on the total number of points in $\bm\Delta_k$, $n(\bm{\Delta}_{k})$. Formally, the joint prior specification for the $K$-set of monotonic functions is given by
\begin{equation}
\pi\left(\lambda_1, \ldots, \lambda_K|\omega, \eta\right) \propto \prod_{k<k'} \exp\Big[-\omega \cdot d_{k,k'} \cdot D_{p,q}\left(\lambda_k, \lambda_{k'}\right)\Big] \times \prod_{k=1}^K\left(1 -\frac{1}{\eta}\right)^{n(\bm{\Delta}_{k})},\omega\geq0,\eta>1.
\label{eq:prior}
\end{equation}
The non-negative constant $d_{k,k'}$ describes our belief on the degree of similarity of regions $k$ and $k'$. Many applications using conditional autoregressive models set $d_{k,k'}=1$ if regions $k$ and $k'$ are adjacent (share a border) and 0 otherwise. Such a choice leads to a decrease in the computational complexity as the integral need not to be evaluated for each pair of functions. This setting is considered in Sections 3 and 4. The parameter $\eta$ in \eqref{eq:prior} refers to the model complexity with the penalty for adding a new point decreasing in $\eta$. Finally, the degree of geographical dependency increases in $\omega$ with $\omega=0$ corresponding to the functions being independent. Hence, the prior $\pi\left(\lambda_1, \ldots, \lambda_K|\omega, \eta\right)$ takes it mode if all functions are equal and constant as $D_{p,q}$, $\omega$, $d_{k,k'}$ and $\eta$ are non-negative.

The probability model \eqref{eq:probmodel} and the constructed joint prior \eqref{eq:prior} fully specify a posterior distribution for the monotonic functions; see Appendix \ref{sec:AppA}. In summary, the posterior distribution of $\lambda_1,\ldots,\lambda_K$ is determined by three components: (a) the specified likelihood in \eqref{eq:probmodel}, (b) the geographical dependency induced by $D_{p,q}$, $\omega$ and $d_{k,k'}$ in \eqref{eq:prior} and (c) the model complexity parameter $\eta$.

\subsection{Inference and Analysis}
\label{sec:Inference}

The statistical framework defined in Subsections \ref{sec:probmodel} to \ref{sec:prior} permits efficient estimation of the underlying regression functions. Inference has to be performed for both the monotonic functions and the smoothing parameter $\omega$. Subsection \ref{sec:Inferencefunctions} outlines the estimation of the monotonic functions $\lambda_1,\ldots,\lambda_K$ while Subsection \ref{sec:BayesOptim} details the estimation of the smoothing parameter $\omega$ via cross-validation and Bayesian global optimization. Finally, Subsection \ref{sec:analysis} considers the analysis of the regression functions $\lambda_1,\ldots,\lambda_K$ based upon the realisations sampled from the posterior.

\subsubsection{Estimation of the monotonic function $\lambda_1,\ldots,\lambda_K$}
\label{sec:Inferencefunctions}

The $K$ monotonic functions are estimated by MCMC techniques, exploiting the marked point process formulation in Section \ref{sec:pointprocess}. Each location $\bm{\xi}_{k,i,j}$ with associated mark $\delta_{k,i,j}$ is considered as one parameter with the number of points, hence the dimension of the parameter space, unknown. All functions are initially constant with predefined level $\delta_{min}$ and each point process $\Delta_{k,i}$ contains no point. Similarly to \citet{Saarela2011}, the marked point processes are then updated sequentially by RJMCMC. More precisely, one of three moves, implying local changes in the regression surface, is proposed on one of the processes $\Delta_{k,1}, \ldots, \Delta_{k,I}$ for region $k$, $k=1,\ldots,K$, in turn. The first move, \textit{Birth}, adds a point $\left(\bm{\xi}^*,\delta^*\right)$ to the process with the level $\delta^*$ being sampled such that monotonicity is preserved. The sample space for $\bm\xi^*$ may, for instance, be $X_k$ or an extended space; the latter facilitating extrapolation of $\lambda_k$. A \textit{Death} removes a point from the current process, maintaining reversibility. Finally, a \textit{Shift} leads to a 'local' change in both the location and level of an existing support point, subject to the monotonic structure of the locations being maintained. For more details on the RJMCMC algorithm see Appendix \ref{sec:AppA}.

In addition to the three moves defined above, the RJMCMC approach requires the specification of a maximum possible number of points, $n_{max}$. However, this does not limit the statistical rigour as the number of points is unlikely to exceed the number of data points. Consequently, integrating \eqref{eq:prior} over the set of potential monotonic functions leads to a finite value due to the boundaries on the covariate spaces, levels (Subsection \ref{sec:probmodel}) and the number of support points. Further, any data based likelihood function implies a proper posterior, even though the prior is improper, in the sense that it has no mean. Therefore, updating the monotonic functions by the proposed RJMCMC algorithm is feasible and approximates, after convergence, the true posterior distribution.

\subsubsection{Estimation of the smoothing parameter $\omega$}
\label{sec:BayesOptim}

Performance of our approach relies on a suitable value for the smoothing parameter $\omega$ in \eqref{eq:prior}. If $\omega$ is too high, the prior dominates the posterior distribution and geographical variation in the regression function is oversmoothed. Otherwise, the data may be overfitted by the estimated function if $\omega$ is too small. The normalising constant of \eqref{eq:prior} can, however, not be calculated analytically. Even though approaches to handle intractable normalising constants are available, these cannot be easily applied in this setting as efficient sampling from the specified prior distribution is hard. Approximate Bayesian computation, for instance, would require sampling multiple times from \eqref{eq:prior} for each update of $\omega$. Hence, here the parameter $\omega$ is not updated while running the RJMCMC algorithm but, rather, is estimated before. 

In this work, a suitable value for $\omega$ is found by $s$-fold cross-validation. The data for each of the $K$ regions are split into $s$ subsets of equal size and the RJMCMC algorithm is performed $s$ times with varying training and test data for each considered value of $\omega$. In order to reduce dependency on the split, multiple repetitions of the $s$-fold cross validation with the same value for $\omega$ are performed. Parameter values are assessed and compared by the mean squared error (MSE) of the posterior predictive functional mean of the test data points derived by Monte Carlo integration. Model comparison may alternatively be considered in terms of the posterior predictive densities. Nevertheless, the number of evaluated values for $\omega$ should be as small as possible since the RJMCMC algorithm described above is computationally expensive. 

We propose to reduce the number of cross-validations by applying Bayesian optimization, in particular, the \textit{efficient global optimization (EGO)} algorithm by \citet{Jones1998}. Despite having potential to reduce the number of evaluations substantially, this concept has, to the best of our knowledge, never been applied in combination with cross-validation. Hence, we outline a new algorithm, termed \textit{EGO-CV}, in the following which combines the two concepts and aims to reduce the computational time. 

The \textit{EGO} concept postulates a sequential design strategy to detect global extrema of black-box functions. \textit{EGO} is widely applied in simulations if the objective function is costly to evaluate and the parameter space is relatively small \citep{Roustant2012}. The rationale is to model the unknown function by a Gaussian process which is updated sequentially after each evaluation and proposals are then based on the expected improvement criterion. More formally, the expected improvement at an arbitrary point $z$, for a Gaussian process $G$, and given the current optimal value, $f_{opt}$, of the unknown function is defined as
\begin{equation}
\mathbb{E}\left[\max\left(f_{opt}-G(z),0\right)\right]
\end{equation}
and represents the potential for $z$ to improve upon the current optimal value. Proposals are considered until the expected improvement falls below a critical value which corresponds to the current solution being close to the unknown optimum. As \textit{EGO} balances between a local exploration of the areas likely to provide 'good model fit' and a global search (in order to avoid a local but not global minima), a suitable parameter value is generally found after a reasonable number of evaluations. 

In the context of estimating $\omega$, interest lies in the global minimum of the unknown cross-validation function, CV($\omega$), and a general layout of our $\textit{EGO-CV}$ approach is given in Algorithm \ref{alg:BSMMR}. Prior to performing Bayesian optimization, an upper bound is derived as \textit{EGO} can only be applied to a closed set. Hence, an initial bound $\omega_u$ is increased until the associated MSE is sufficiently greater than the one obtained for $\omega=0$. More clarity is provided in lines 2 to 7 in Algorithm \ref{alg:BSMMR}. In this paper, an upper bound based upon $\beta=1.1$ in Algorithm \ref{alg:BSMMR} proved reasonable in all simulations. Once the bound is fixed, an initial proposal $\omega^*\in\left[0, \omega_u\right]$ is made, guaranteeing that the Gaussian process is fitted with at least 3 data points. After performing cross-validation for $\omega^*$, the \textit{EGO} algorithm is performed until the expected improvement falls below the critical value $\alpha$. The value $\omega_{opt}$ providing the lowest MSE is finally used as the smoothing parameter $\omega$ in the conclusive RJMCMC algorithm. In this work, \textit{EGO} is performed by the \texttt{DiceOptim} package implemented in \texttt{R} by \citet{Roustant2012}. In addition to the MSE, its variance across the repetitions is derived too, as the \texttt{DiceOptim} package allows to account for uncertainty in the function evaluations. The parameter $\eta$ may be estimated similarly by investigating the regions separately or setting $\omega=0$. The simulations in Section 3 focus, however, solely on the estimation of $\omega$.

\begin{algorithm}
\caption{\textit{Efficient Global Optimization within Cross-Validation (EGO-CV)}}
\label{alg:BSMMR}
\begin{algorithmic}[1]
\REQUIRE {Observations and parameter settings for RJMCMC algorithm}
\REQUIRE {Cross-validation parameters: number of folds and number of repetitions}
\REQUIRE {Initial upper bound $\omega_u$, critical value $\alpha$, factor $\beta$}
\STATE Initialize cross validation results \texttt{cv\_MSE} and expected improvement \texttt{max\_EI}$>\alpha$
\STATE Perform cross-validation for $\omega=0$ and store CV(0) in \texttt{cv\_MSE}
\STATE Perform cross-validation for $\omega_u$ and store CV($\omega_u$) in \texttt{cv\_MSE}
\WHILE {CV($\omega_u$) $<$ $\beta$ CV(0)}
\STATE Increase upper bound $\omega_u$
\STATE Perform cross-validation for new $\omega_u$ and store CV($\omega_u$)
\ENDWHILE
\STATE Set initial proposal $\omega^*$, e.g.\ $\omega^* = \omega_u/2$ 
\WHILE {\texttt{max\_EI} $>\alpha$}
\STATE Perform cross-validation for $\omega^*$ and store CV($\omega^*$) in \texttt{cv\_MSE}
\STATE Perform \textit{EGO} on the interval $[0,\omega_u]$ and update $\omega^*$ and \texttt{max\_EI}  
\ENDWHILE
\RETURN {Parameter value $\omega_{opt}$ which provides smallest mean squared error in \texttt{cv\_MSE}}
\end{algorithmic}
\end{algorithm} 

\subsubsection{Analysis}
\label{sec:analysis}

The RJMCMC algorithm in Subsection \ref{sec:Inferencefunctions} and Appendix \ref{sec:AppA} runs for a fixed number of iterations and thinning is performed in order to reduce autocorrelation of the samples which is high as the functions change locally only. Convergence is checked by sampling uniformly a fixed number of points from the covariate space and investigating the associated trace plots and auto-correlation functions.

Realisations sampled from the posterior distribution are rich and facilitate detailed analysis of the estimated monotonic functions. Posterior estimates for $\lambda_k$ are obtained by averaging over stored realisations sampled in the Bayesian framework. Both smooth and discontinuous functional forms can be recovered by averaging over a large number of realisations with varying number, locations and levels of the points \citep{Heikkinen2003}. The posterior mean and quantiles are accessible for any covariate value $\mathbf{x}_k\in X_k$ by deriving the associated functional level $\lambda_k^{(r)}(\mathbf{x}_k)$ for each sample $r,~r=1,\ldots,R$. Plots of the posterior functional mean are obtained by evaluating the estimated posterior mean for a finite set of covariate values, e.g.\ by defining a regular grid on $X_k$.

Finally, the samples also facilitate the detection of threshold effects in the regression surface. This detection requires distinguishing the points in the sampled marked point processes into those representing a threshold effect and those approximating a continuous shape. In general, threshold effects are expected to occur in most of the samples, i.e.\ they are removed with low probability and a shift is only likely to be accepted if it changes the point marginally. Here, threshold effects are additionally defined in terms of representing a large functional level change in the regression surface. Based on these considerations, each sampled point is classified. Points across samples are considered as the same threshold effect if both points are very close in the covariate space and have similar functional levels. Based on this classification, potential threshold effects are listed and their empirical occurrence rate across samples is derived.
\section{Simulation Study}
\label{sec:Sim}

This section aims to demonstrate that BSMMR is highly flexible, in terms of reconstructing a wide range of regression surfaces, and appraise the value for sharing statistical information geographically between regions. Multiple simulations studies are performed in order to
\begin{enumerate}
\item Illustrate that BSMMR improves estimates if similarities between functional shapes exist, and is also robust if the functions are dissimilar
\item Verify that \textit{EGO-CV}, Algorithm \ref{alg:BSMMR} in Section \ref{sec:BayesOptim}, yields a sensible value for $\omega$ 
\item Examine sensitivity on the prior parameters $p$, $q$ and $\eta$ in \eqref{eq:prior}.
\end{enumerate}
The first two goals are considered by comparing results to those for $\omega=0$ which corresponds to imposing no geographical dependency between functions. In order to facilitate visualisation of the estimated posterior mean functions, covariate spaces are bivariate in all simulations. Estimates are assessed by the mean absolute error (MAE) and the standard deviation of the difference between posterior mean and true underlying function on a regular $100\times100$ grid of the covariate space. 

The \textit{EGO-CV} algorithm is applied with 5 repetitions of a 10-fold cross-validation. Estimates for each fold are obtained by performing 50,000 iterations and storing every 100th draw after a burn-in of 25,000. The initial bound $\omega_u$ is set to 50 and increased by factor 10 until CV($\omega_u$) is at least 10\% higher than CV(0). The critical value $\alpha$ is set to 0.01\% of the current minimum, i.e.\ $\alpha = \min($\texttt{cv\_MSE}$)/10000$. Additionally, the algorithm also stops if 30 values for $\omega$ have been considered. Since smoothing is more sensitive on lower than upper values for $\omega$, \textit{EGO} is performed on a transformed scale with $\widetilde{\omega} = \sqrt{\omega/50}$ which provided increased robustness. Alternatively, \textit{EGO} may also be applied on a transformed log scale, etc.

The RJMCMC algorithm described in Section \ref{sec:Inference} and Appendix \ref{sec:AppA} then runs with the derived parameter value $\omega_{opt}$ for 2,500,000 iteration steps after a burn-in period of 500,000 and every 1000th iteration is considered for analysis. The maximum number of points, $n_{max}$, is fixed to 200 and \textit{Birth}, \textit{Death} and \textit{Shift} are proposed with probabilities 0.3, 0.3 and 0.4, respectively. Convergence of the algorithm is checked by investigating the trace plots of the functional levels for ten random points for each region $k$, $k=1,\ldots,K$; examples are provided in \citet{Rohrbeck2016}. 

Section \ref{sec:SimGauss} considers two contiguous regions with Gaussian response data and illustrates sensitivity analysis on the model complexity parameter $\eta$ and the prior parameters $p$ and $q$ in Subsections \ref{sec:SimGausseta} and \ref{sec:SimGausspq}, respectively. Subsection \ref{sec:SimGausseta} also compares BSMMR to a geographically varying coefficient (GVC) model with an unique CAR prior on each covariate effect. More complex geographical networks with Binomial response data and varying covariate spaces are considered in Section \ref{sec:SimBinom}.

\subsection{Gaussian Data}
\label{sec:SimGauss}

Observations for region $k=1,2$ are simulated independently from a Normal distribution 
\begin{equation}
y_k\sim\mathcal{N}\left(~\alpha_k + \lambda_k(\mathbf{x}_k),~ \sigma_k^2~\right), 
\label{eq:SimGauss}
\end{equation}
with the monotonic functions $\lambda_1$ and $\lambda_2$ both defined on the unit square: $X_1=X_2=[0,1]^2$. The distribution of the covariate values varies across the two sets of simulations and is described in the respective subsections. Functional levels $\alpha_k + \lambda_k(\mathbf{x}_k)$ lie between 0 and 2 across simulations, facilitating comparability of the different settings. The variances $\sigma_k^2$, $k=1,2$, are treated as unknown, with Inverse-Gamma priors, and are updated by Gibbs sampling. In Subsection \ref{sec:SimGausseta}, a CAR prior \citep {Besag1991} is placed on the the baseline levels $\alpha_k$, $k=1,2$, and these are estimated using Metropolis-within-Gibbs. For comparison they are fixed to $\alpha_k=0$ in Subsection \ref{sec:SimGausspq}. 
 
\subsubsection{Sensitivity analysis for model complexity $\eta$}
\label{sec:SimGausseta}

\begin{figure}
\begin{center}
\parbox{2.8cm}{\hfill}
\parbox{4.8cm}{\begin{center}\hspace{0.2cm}\textbf{Region 1 - Truth}\end{center}}
\parbox{4.8cm}{\begin{center}\hspace{0.2cm}\textbf{Region 2 - Truth}\end{center}}
\parbox{4.8cm}{\begin{center}\hspace{0.2cm}\textbf{Region 2 - $\omega=\omega_{opt}$}\end{center}}\\      
\vspace{-0.1cm}
\parbox{2.8cm}{\textbf{Study 1}\\ \vspace{0cm}\\Identical:\\ Continuous}  
\begin{minipage}{4.8cm}\begin{center} \includegraphics[width=4.15cm]{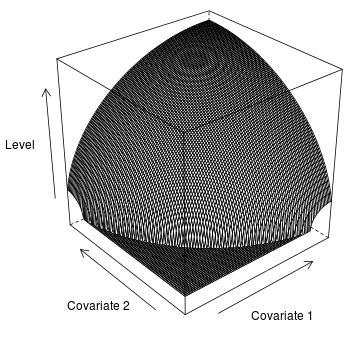}\end{center}\end{minipage} 
\begin{minipage}{4.8cm}\begin{center} \includegraphics[width=4.15cm]{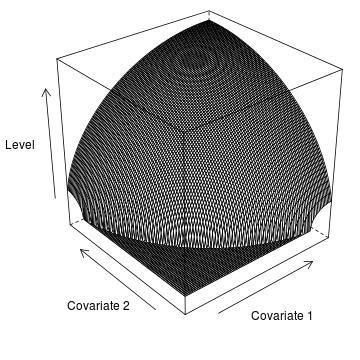}\end{center}\end{minipage}
\begin{minipage}{4.8cm}\begin{center} \includegraphics[width=4.15cm]{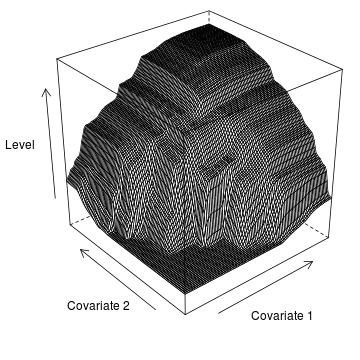}\end{center}\end{minipage}\\
  
\parbox{2.8cm}{\textbf{Study 2}\\ \vspace{0cm}\\Identical:\\ Discontinuous} 
\begin{minipage}{4.8cm}\begin{center} \includegraphics[width=4.15cm]{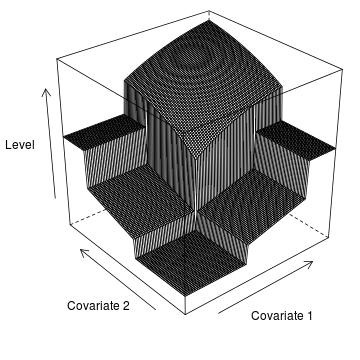}\end{center}\end{minipage} 
\begin{minipage}{4.8cm}\begin{center} \includegraphics[width=4.15cm]{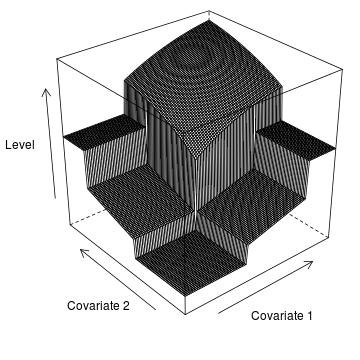}\end{center}\end{minipage}
\begin{minipage}{4.8cm}\begin{center} \includegraphics[width=4.15cm]{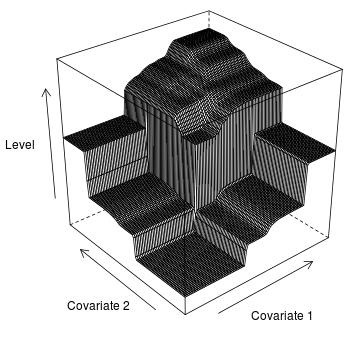}\end{center}\end{minipage}\\
  
\parbox{2.8cm}{\textbf{Study 3}\\ \vspace{0cm}\\Similar:\\ Continuous} 
\begin{minipage}{4.8cm}\begin{center} \includegraphics[width=4.15cm]{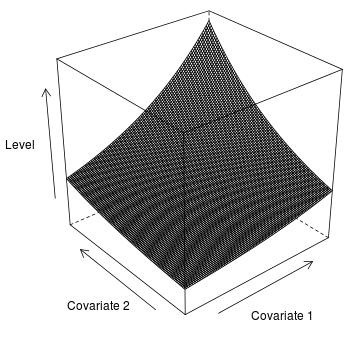}\end{center}\end{minipage}  
\begin{minipage}{4.8cm}\begin{center} \includegraphics[width=4.15cm]{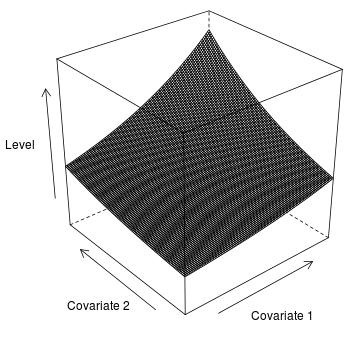}\end{center}\end{minipage}
\begin{minipage}{4.8cm}\begin{center} \includegraphics[width=4.15cm]{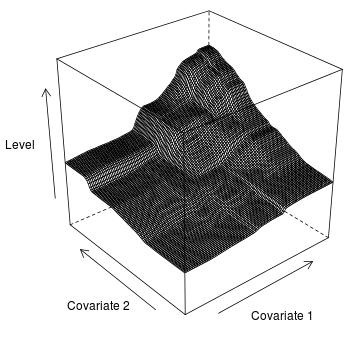}\end{center}\end{minipage}\\
    
\parbox{2.8cm}{\textbf{Study 4}\\ \vspace{0cm}\\Similar:\\ Discontinuous} 
\begin{minipage}{4.8cm}\begin{center} \includegraphics[width=4.15cm]{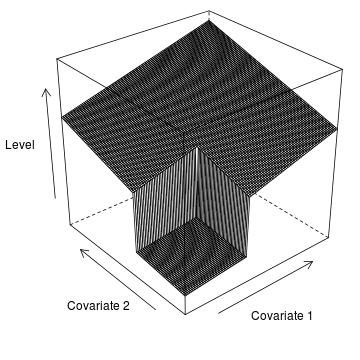}\end{center}\end{minipage} 
\begin{minipage}{4.8cm}\begin{center} \includegraphics[width=4.15cm]{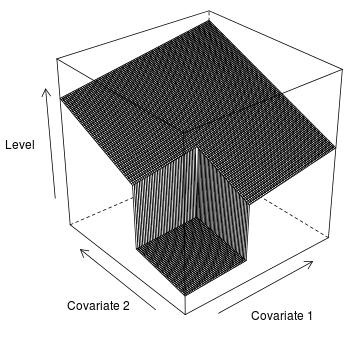}\end{center}\end{minipage}
\begin{minipage}{4.8cm}\begin{center} \includegraphics[width=4.15cm]{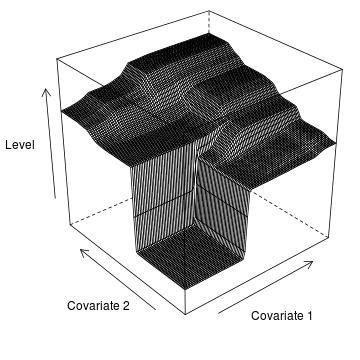}\end{center}\end{minipage}\\
    
\parbox{2.8cm}{\textbf{Study 5}\\ \vspace{0cm}\\Different:\\ Continuous} 
\begin{minipage}{4.8cm}\begin{center} \includegraphics[width=4.15cm]{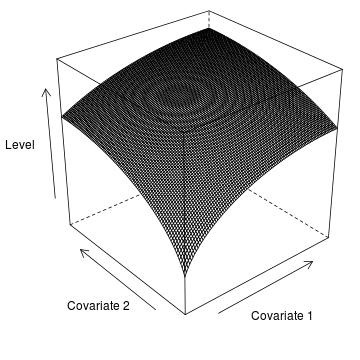} \end{center}\end{minipage} 
\begin{minipage}{4.8cm}\begin{center} \includegraphics[width=4.15cm]{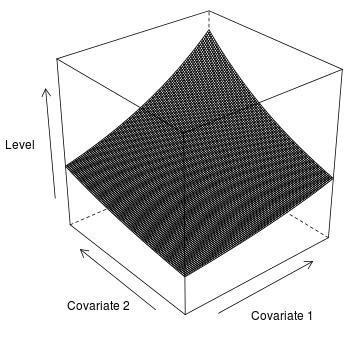} \end{center}\end{minipage}
\begin{minipage}{4.8cm}\begin{center} \includegraphics[width=4.15cm]{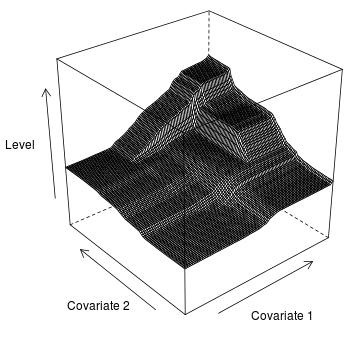}  \end{center}\end{minipage}
           
\caption{True functions for region 1 (first column) and region 2 (second column), and the posterior mean for region 2 obtained by BSMMR with $\omega=\omega_{opt}$ and $\eta=10$ for the five pairs of functions in Subsection \ref{sec:SimGausseta}.}
\label{fig:Sim31}
\end{center}
\end{figure}

To explore the flexibility of BSMMR to fit a wide range of functional shapes, various pairs of monotonic functions are considered. The true underlying functions $\lambda_1$ and $\lambda_2$, illustrated in the first two columns of Figure \ref{fig:Sim31}, range from smooth curves through to discontinuous surfaces with several threshold effects. Data are sampled such that 1,000 data points are observed for region 1 while 100 are observed for region 2. Hence, $\lambda_1$ can be estimated more precisely than $\lambda_2$. This scenario facilitates, in particular, examination of the potential benefits of estimating $\lambda_2$ when borrowing statistical information from region 1. Covariate values are sampled uniformly across the unit square, $\mathbf{x}_k \sim \mathcal{U}([0,1]^2)$, and $\sigma_k^2=0.05^2$, $k=1,2$. Three values are considered for the model complexity parameter, $\eta=(2,10,1000)$, corresponding respectively, to high, moderate and low penalties for adding a new point. The remaining parameters in the prior density in \eqref{eq:prior} are set to $p=q=1$, so that $D_{p,q}$ simplifies to the integrated squared difference. The considered GVC model consists of one intercept and two covariate effects for region $k=1,2$ and parameters are estimated by a Metropolis-within-Gibbs algorithm with 100,000 iterations.

\begin{table}
\begin{center}
\caption{Mean absolute errors$\times 10^{-2}$ and standard deviations$\times 10^{-2}$ of the difference between the true function and posterior mean for different values of the model complexity parameter $\eta$ for the five considered pairs of monotonic functions in Studies 1 to 5. The first row for each study refers to region 1. The first column gives the results for performing BSMMR with the derived parameter value $\omega_{opt}$ while the second column gives the results for $\omega=0$. The last column contains the results for an estimated GVC model.}
\label{tab:Sim31}
\begin{tabular}{cccccccc}
\hline
\Tstrut & \multicolumn{2}{c}{$\eta=2$} & \multicolumn{2}{c}{$\eta=10$} & \multicolumn{2}{c}{$\eta=1000$}              & GVC         \Bstrut\\
Study & $\omega = \omega_{opt}$ & $\omega=0$ & $\omega = \omega_{opt}$ & $\omega=0$ & $\omega = \omega_{opt}$ & $\omega=0$  &             \Bstrut\\ 
\hline    
\textbf{1} & 3.4 (5.2)               & 3.3 (5.2)  & 3.3 (5.1)               & 3.1 (4.9)  & 3.2 (4.9)               & 3.3 (5.3)   & 15.3 (18.5) \Tstrut\\
           & 4.9 (6.6)               & 7.1 (9.7)  & 4.7 (6.2)               & 6.4 (8.8)  & 4.8 (6.4)               & 6.6 (8.8)   & 15.2 (18.5) \Bstrut\\

\textbf{2} & 2.0 (4.0)               & 1.9 (3.5)  & 1.9 (3.5)               & 1.9 (3.6)  & 1.8 (3.4)               & 2.0 (4.0)   & 14.9 (18.0) \Tstrut\\
           & 3.1 (4.5)               & 4.2 (7.0)  & 2.8 (4.2)               & 4.3 (6.9)  & 2.9 (4.6)               & 4.2 (6.7)   & 14.9 (18.0) \Bstrut\\

\textbf{3} & 2.7 (3.5)               & 2.8 (3.7)  & 2.6 (3.4)               & 2.6 (3.4)  & 2.5 (3.3)               & 2.4 (3.2)   & 6.9 (8.7)   \Tstrut\\
           & 3.5 (4.7)               & 4.2 (5.5)  & 3.0 (4.1)               & 3.9 (5.2)  & 2.7 (3.6)               & 3.8 (4.9)   & 5.3 (7.1)   \Bstrut\\

\textbf{4} & 2.7 (4.7)               & 2.7 (5.1)  & 2.7 (5.0)               & 2.6 (4.7)  & 2.6 (5.1)               & 2.6 (4.5)   & 13.8 (17.8) \Tstrut\\
           & 4.5 (6.4)               & 5.8 (11.6) & 4.3 (6.1)               & 5.8 (11.5) & 4.6 (6.4)               & 5.5 (11.3)  & 14.8 (18.8) \Bstrut\\

\textbf{5} & 2.2 (2.8)               & 2.3 (2.9)  & 2.2 (2.8)               & 2.2 (2.8)  & 2.2 (2.8)               & 2.2 (2.8)   & 4.9 (6.4)   \Tstrut\\
           & 4.9 (6.5)               & 5.1 (6.5)  & 4.7 (6.0)               & 4.7 (6.0)  & 4.8 (6.1)               & 4.8 (6.1)   & 4.7 (6.9)   \Bstrut\\
\hline
\end{tabular}
\end{center}
\end{table}

Study 1 and 2 in Figure \ref{fig:Sim31} and Table \ref{tab:Sim31} consider cases where the regional functions are identical: $\lambda_1 = \lambda_2$. Both the MAE and the standard deviation of the difference decrease, in particular, for region 2 by performing BSMMR with $\omega_{opt}$. The estimated posterior mean for $\lambda_2$ in the final column of Figure \ref{fig:Sim31} illustrates that both the smooth surfaces and threshold effects are captured well. The estimated GVC models perform worse due to the non-linearity of the true underlying functions. Study 3 and 4 consider cases where $\lambda_1$ and $\lambda_2$ are similar and the conclusions align with those for Study 1 and 2. The improvement is also visible in the posterior mean plots for $\omega=0$ and $\omega=\omega_{opt}$ in Study 1 and 4, see Appendix \ref{sec:AppB} for details. With respect to $\eta$, the results show only slight differences in the model fit. One exception is that the model fit in Study 3 improves consistently with increasing $\eta$. Finally, Study 5 applies the algorithm to a case where the functions are different. The results show no worsening for both region 1 and region 2 as the estimated smoothing parameter $\omega_{opt}$ is indeed equal $0$ for $\eta=10$ and $\eta=1000$. Posterior mean plots for all settings are provided in \citet{Rohrbeck2016}.

The five simulation studies illustrate that BSMMR may be used effectively to improve estimates by borrowing statistical information from adjacent regions. Results for region 2 improve for all cases with similar shapes which indicates that BSMMR is able to exploit neighbourhood information regardless of the underlying functions. The proposed \textit{EGO-CV} algorithm returns a suitable and robust value $\omega_{opt}$ which does not oversmooth even if the functional shapes show no similarities. Slight, or no, variations are found with respect to $\eta$, with only Study 3 showing a consistent improvement of the model fit with increasing $\eta$. As higher values for $\eta$ allow on average for a higher number of process points, the smooth surfaces are fitted better due to the posterior mean having more but smaller jumps. Since the simulations indicate little, or no, sensitivity, the parameter value is fixed to $\eta=2$ in the following simulations.

\subsubsection{Sensitivity analysis for prior parameters $p$ and $q$}
\label{sec:SimGausspq} 

Subsection \ref{sec:SimGausseta} explored the performance of BSMMR for different functional shapes but only considered uniformly distributed covariate values. In the following simulations, interest lies in exploring the flexibility of BSMMR for handling non-uniform distributions of $\mathbf{x}$. This setting also allows a more general sensitivity analysis on $p$ and $q$. In particular, performance may depend on whether relatively more data points are observed in areas with similar functional levels. 

\begin{figure}[ht]
\begin{center}
\parbox{4.3cm}{\begin{center} \hspace{0.3cm}\textbf{Truth}   \end{center}}
\parbox{4.3cm}{\begin{center} \hspace{0.3cm}\textbf{Study 1} \end{center}}
\parbox{4.3cm}{\begin{center} \hspace{0.3cm}\textbf{Study 2} \end{center}} 
\parbox{4.3cm}{\begin{center} \hspace{0.3cm}\textbf{Study 3} \end{center}}\\
\begin{minipage}{4.3cm}\begin{center}\includegraphics[width=4.3cm]{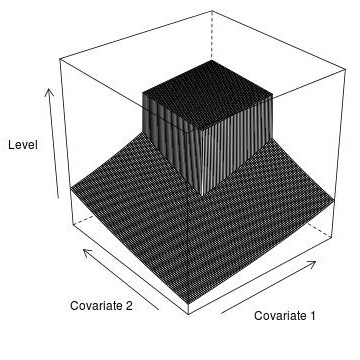}\end{center} \end{minipage}
\begin{minipage}{4.3cm}\begin{center}\includegraphics[width=4.3cm]{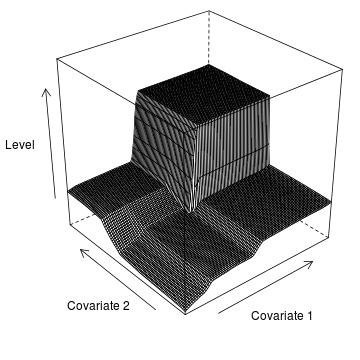}\end{center} \end{minipage}
\begin{minipage}{4.3cm}\begin{center}\includegraphics[width=4.3cm]{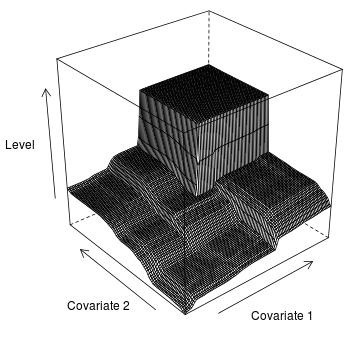}\end{center} \end{minipage}
\begin{minipage}{4.3cm}\begin{center}\includegraphics[width=4.3cm]{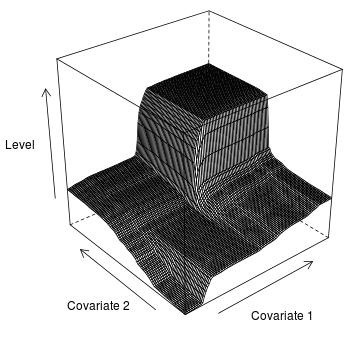}\end{center} \end{minipage}\\
\begin{minipage}{4.3cm}\begin{center}\includegraphics[width=4.3cm]{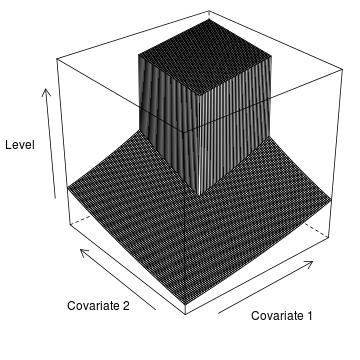}\end{center} \end{minipage} 
\begin{minipage}{4.3cm}\begin{center}\includegraphics[width=4.3cm]{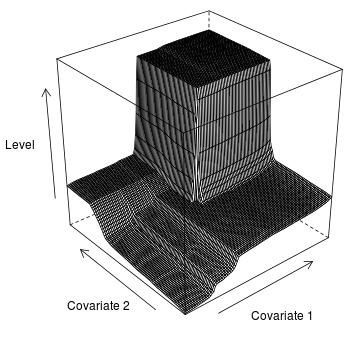}\end{center} \end{minipage}
\begin{minipage}{4.3cm}\begin{center}\includegraphics[width=4.3cm]{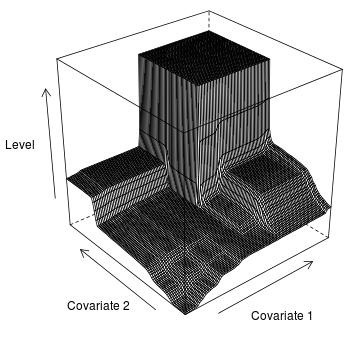}\end{center} \end{minipage}
\begin{minipage}{4.3cm}\begin{center}\includegraphics[width=4.3cm]{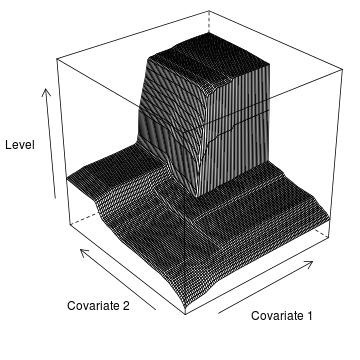}\end{center} \end{minipage}
\caption{True functions and posterior mean plots for the parameter settings of $p$ and $q$ providing the lowest combined MAE by performing BSMMR with the proposed value $\omega_{opt}$
for Study 1 to 3. Plots for region 1 are given in the first row while the second row refers to region 2. The threshold effect is at $(0.5, 0.5)$ in the true functions. The 200 covariate values for each region are sampled uniformly on $[0,1]^2$ in Study 2 and in the ratios $150:50$ and $25:175$, above:below the threshold, respectively, in Studies 1 and 3.}
\label{fig:Sim32}
\end{center}
\end{figure}

The first column in Figure \ref{fig:Sim32} illustrates the true pair of underlying monotonic functions which is fixed across Studies 1 to 3. Both functions exhibit a threshold effect at $(0.5, 0.5)$ and have similar lower functional levels. For each of the two regions, 200 data points are simulated with $\sigma_1^2=0.2^2$ and $\sigma_2^2=0.3^2$. The number of data points sampled above the threshold effect changes across simulations. Study 1 considers the case with 150 covariate values sampled uniformly above the threshold effect $(0.5, 0.5)$. Study 3 considers the case where 25 covariate values are sampled for the upper functional levels. The remaining data points, 50 and 175, respectively, are sampled uniformly below the threshold effect. Study 2 considers the case with the covariate values being sampled uniformly from the unit square, $\mathbf{x}\sim\mathcal{U}\left[0,1\right]^2$.

Five different settings for $p$ and $q$ are applied and compared. The first two settings: (1) $p=1,q=1$ and (2) $p=1,q=2$ impose a constant degree of dependency between $\lambda_1$ and $\lambda_2$, independent of the functional levels. Settings (3) $p=0.5,q=1$ and (4) $p=-1,q=1$, allow for stronger dependency in the lower functional levels while (5) $p=3,q=1$ imposes increased dependency for higher levels. 

Table \ref{tab:Sim32} shows that the settings with $p<1$ perform generally best and improve the average MAE by up to $17\%$ compared to $\omega=0$. The degree of improvement by sharing information geographically decreases from Study 1 to 3 due to the number of data points being available to fit the lower functional levels. Figure \ref{fig:Sim32} illustrates that the threshold effect is captured correctly in each study. Finally, the model fit for region 1 is better than for region 2 across simulations due to the reduced variability in the observations. Posterior mean plots for all settings are provided in \citet{Rohrbeck2016}

In Study 1, all settings for $p$ and $q$ improve the model fit in terms of the averaged MAE and standard deviation compared to the setting $\omega=0$. The overall improvement is due to the higher concentration of data points above the threshold. In particular, a value $\omega_{opt}$ is found such that the prior contributes beneficially to the estimation of the lower functional levels without causing a large bias on the upper functional levels. The settings with $p<1$ perform best as these impose a very small penalty for differences in the upper levels, hence allowing for increased dependency in the lower functional levels. The posterior mean plots in the second column of Figure \ref{fig:Sim32} illustrate that some features in the lower functional levels are not captured well due to the low number of data points available. 

For Study 2, Table \ref{tab:Sim32} shows that the settings $p\geq1$ perform similarly to $\omega=0$ whilst $p<1$ leads to substantial improvements for uniformly distributed data. As the intensity of data points is similar across the covariate space, values for $\omega$ leading to improvements for the lower functional levels affect the estimation of the upper functional levels more strongly than in Study 1. As such, settings with $p\geq1$ cause some bias in the upper functional levels while improving estimates for the lower functional levels. Figure \ref{fig:Sim32} also shows that the lower functional levels are fitted better when compared to Study 1. Finally, all settings except for $p=-1,q=1$ perform similarly to $\omega=0$ in Study 3. As the intensity of observations is increased on lower levels, compared to upper levels, $\omega_{opt}$ often implies a bias for the upper levels whilst improving model fit on the lower levels. This does not occur for $p=-1, q=1$ as dependency decreases relatively quickly with increasing functional level.

In summary, the simulations performed in this subsection illustrate two important aspects of the BSMMR approach. Firstly, the chosen values for $p$ and $q$ affect the performance of BSMMR quite strongly. As such, $p$ should be chosen to be smaller than 1 if functions are presumably similar in their lower functional levels only. Conversely, $p$ should be set to be greater than 1 if the upper levels are more similar. Also, the appropriateness of the choice for $p$ and $q$ depends not only on functional similarities but also on the distribution of the covariate values.   

\begin{table}
\begin{center}
\caption{Mean absolute errors$\times 10^{-2}$ and standard deviation$\times 10^{-2}$ of the difference between true function and posterior mean for various settings of the prior parameters $p$ and $q$. The first row for each study refers to region 1 and the last column gives the results for $\omega=0$.}
\label{tab:Sim32}
\begin{tabular}{ccccccc}
\hline
\Tstrut Study  & $p=1,q=1$   & $p=1,q=2$  & $p=0.5,q=1$           & $p=-1,q=1$             & $p=3,q=1$   & $\omega=0$  \Bstrut\\
\hline
\textbf{1}	   & 6.3 (9.7)   & 6.7 (11.0) & 6.2 (8.9)             & $\mathbf{6.6~ (10.3)}$ & 5.7 (7.2)   & 6.2 (9.3)   \Tstrut\\
 				       & 8.9 (12.1)  & 9.1 (12.7) & 8.3 (11.5)            & $\mathbf{7.4~ (10.8)}$ & 9.2 (12.7)  & 10.6 (15.0) \Bstrut\\

\textbf{2}	   & 5.4 (7.5)   & 5.3 (7.3)  & $\mathbf{5.0~ (6.7)}$ & 5.1 (7.1)              & 5.2 (7.0)   & 4.8 (6.5)   \Tstrut\\
 				       & 7.4 (10.2)  & 7.2 (9.3)  & $\mathbf{6.8~ (9.2)}$ & 7.2 (9.6)              & 7.6 (10.0)  & 8.0 (10.5)  \Bstrut\\

\textbf{3}	   & 5.6 (7.4)   & 5.7 (7.1)  & 5.5 (7.7)             & $\mathbf{5.1~ (7.7)}$  & 5.5 (8.3)   & 5.9 (8.5)   \Tstrut\\
 				       & 11.1 (14.3) & 10.3(12.7) & 10.5 (14.9)           & $\mathbf{9.3~ (12.6)}$ & 10.4 (15.3) & 10.6 (17.5) \Bstrut\\
\hline
\end{tabular}
\end{center}
\end{table}

\subsection{Binomial Data}
\label{sec:SimBinom}

In this subsection, BSMMR is applied to two geographical networks of 5 regions in order to explore its performance in more complex settings. Observations of the response variable are taken to be Binomially distributed, $y_k\sim\mbox{Binomial}(A, p_k)$, with the number of trials $A=100$ fixed and the success probability for region $k$, $p_k$, on the logit scale being modelled by a monotonic function $\lambda_k$. Formally, responses $y_k$ are simulated from  
\begin{equation}
y_k \sim \mbox{Binomial}\left(100, \frac{\exp{\left[\lambda_k\left(\mathbf{x}_k\right)\right]}}{1+\exp{\left[\lambda_k\left(\mathbf{x}_k\right)\right]}}\right), ~k=1,\ldots,5,
\end{equation}  
where $\lambda_k\left(\cdot\right)$ takes values between 0 and 3, i.e.\ $p_k$ is assumed to lie between 0.5 and 0.95. In the first simulation study, covariate values $\mathbf{x}_k$ are sampled uniformly on $[0,1]^2$. In the second study, the covariate spaces $X_1,\ldots,X_5$ vary. The variation of the covariate spaces is considered to facilitate analysis with respect to $A_{k,k'}$ in \eqref{eq:D}.

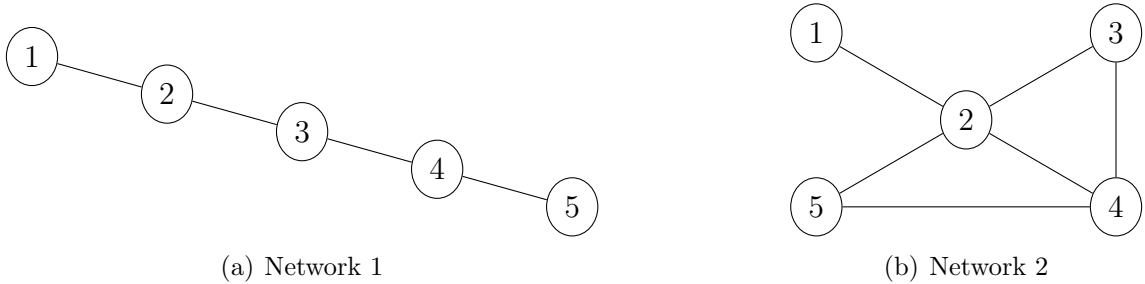
\begin{figure}
\begin{center}
\subfigure[Network 1]{
\label{fig:Sim33a1}
\begin{tikzpicture}[node distance=1.1cm and 1.1cm, mynode/.style={draw,circle,text width=1.5cm,align=center}]
\node[draw, ellipse] (Region1) {1};
\node[draw, ellipse, right=of Region1, yshift=-0.5cm] (Region2){2};
\node[draw, ellipse, right=of Region2, yshift=-0.5cm] (Region3){3};
\node[draw, ellipse, right=of Region3, yshift=-0.5cm] (Region4){4};
\node[draw, ellipse, right=of Region4, yshift=-0.5cm] (Region5){5};
\draw(Region1) to (Region2);
\draw(Region2) to (Region3);
\draw(Region3) to (Region4);
\draw(Region4) to (Region5);
\end{tikzpicture}
}
\hspace{2cm}
\subfigure[Network 2]{
\label{fig:Sim33a2}
\begin{tikzpicture}[node distance=1cm and 1cm, mynode/.style={draw,circle,text width=1.3cm,align=center}]
\node[draw, ellipse] (Region2) {2};
\node[draw, ellipse, above left=of Region2, yshift=-0.4cm, xshift=-0.5cm] (Region1){1};
\node[draw, ellipse, above right=of Region2, yshift=-0.4cm, xshift=0.5cm] (Region3){3};
\node[draw, ellipse, below right=of Region2, yshift=0.4cm, xshift=0.5cm] (Region4){4};
\node[draw, ellipse, below left=of Region2, yshift=0.4cm, xshift=-0.5cm] (Region5){5};
\draw(Region1) to (Region2);
\draw(Region2) to (Region3);
\draw(Region3) to (Region4);
\draw(Region4) to (Region5);
\draw(Region2) to (Region5);
\draw(Region2) to (Region4);
\end{tikzpicture}
}
\caption{Geographical networks of the two simulation studies performed for Binomial data.} 
\label{Fig:Sim33a}
\end{center}
\end{figure}

\begin{figure}
\begin{center}
\parbox{2.8cm}{\hfill} 
\parbox{4.8cm}{\begin{center}\hspace{0.2cm}\textbf{Truth}      \end{center}} 
\parbox{4.8cm}{\begin{center}\hspace{0.2cm}\textbf{$\omega=0$} \end{center}} 
\parbox{4.8cm}{\begin{center}\hspace{0.2cm}\textbf{$p=-1,~q=1$}\end{center}}\\      
\vspace{-0.1cm}
\parbox{2.8cm}{\textbf{Region 1} }  
\begin{minipage}{4.8cm}\begin{center} \includegraphics[width=4.15cm]{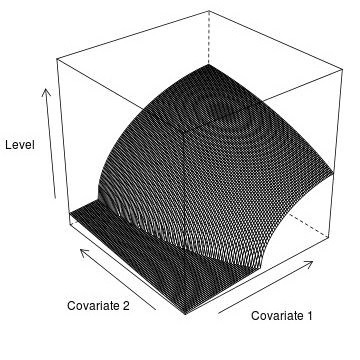} \end{center}\end{minipage} 
\begin{minipage}{4.8cm}\begin{center} \includegraphics[width=4.15cm]{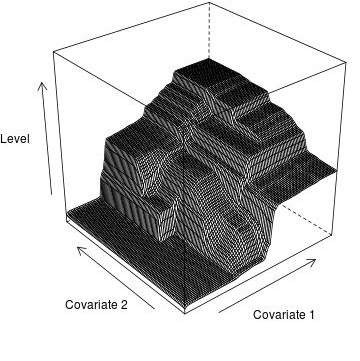} \end{center}\end{minipage}
\begin{minipage}{4.8cm}\begin{center} \includegraphics[width=4.15cm]{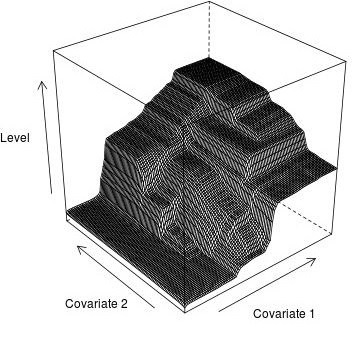}\end{center}\end{minipage}\\
  
\parbox{2.8cm}{\textbf{Region 2} } 
\begin{minipage}{4.8cm}\begin{center} \includegraphics[width=4.15cm]{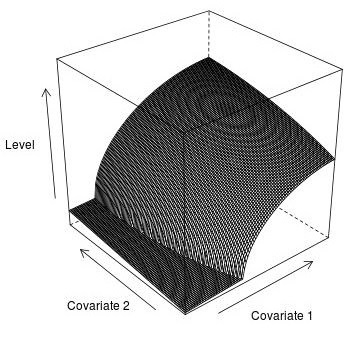} \end{center}\end{minipage} 
\begin{minipage}{4.8cm}\begin{center} \includegraphics[width=4.15cm]{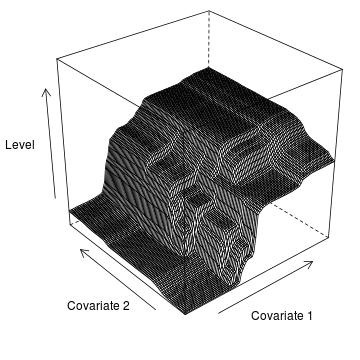} \end{center}\end{minipage}
\begin{minipage}{4.8cm}\begin{center} \includegraphics[width=4.15cm]{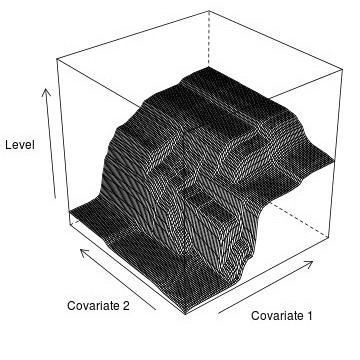}\end{center}\end{minipage}\\
  
\parbox{2.8cm}{\textbf{Region 3} } 
\begin{minipage}{4.8cm}\begin{center} \includegraphics[width=4.15cm]{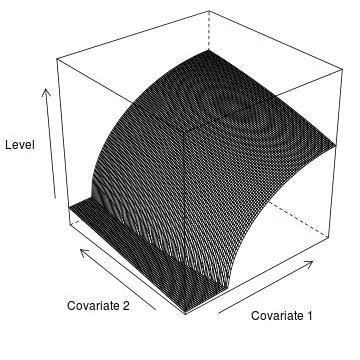}\end{center}\end{minipage}  
\begin{minipage}{4.8cm}\begin{center} \includegraphics[width=4.15cm]{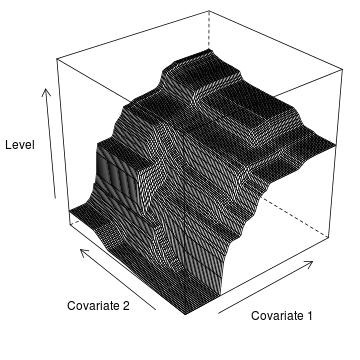} \end{center}\end{minipage}
\begin{minipage}{4.8cm}\begin{center} \includegraphics[width=4.15cm]{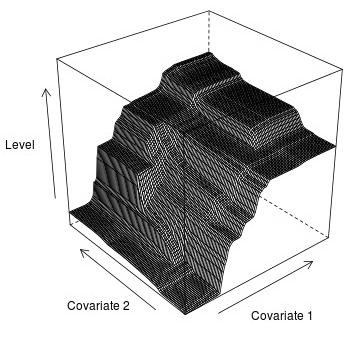}\end{center}\end{minipage}\\
    
\parbox{2.8cm}{\textbf{Region 4} } 
\begin{minipage}{4.8cm}\begin{center} \includegraphics[width=4.15cm]{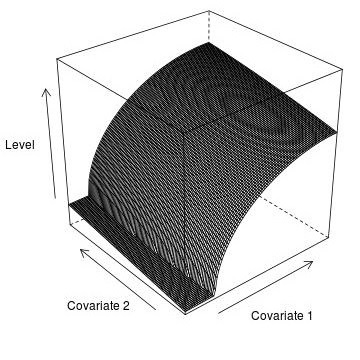}\end{center}\end{minipage} 
\begin{minipage}{4.8cm}\begin{center} \includegraphics[width=4.15cm]{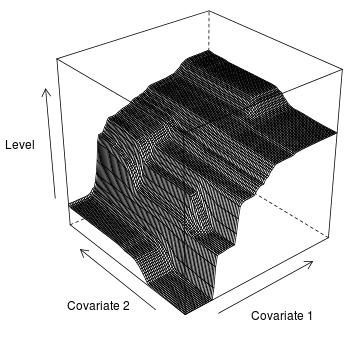} \end{center}\end{minipage}
\begin{minipage}{4.8cm}\begin{center} \includegraphics[width=4.15cm]{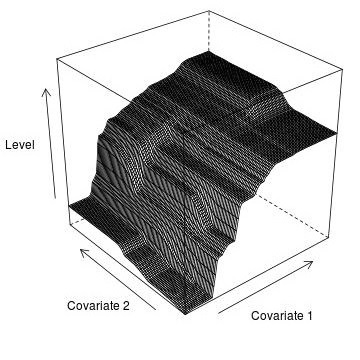}\end{center}\end{minipage}\\
    
\parbox{2.8cm}{\textbf{Region 5} } 
\begin{minipage}{4.8cm}\begin{center} \includegraphics[width=4.15cm]{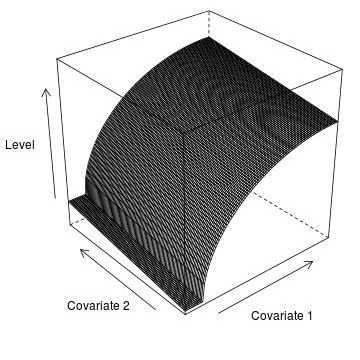}\end{center}\end{minipage} 
\begin{minipage}{4.8cm}\begin{center} \includegraphics[width=4.15cm]{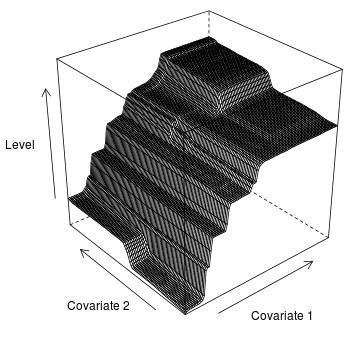}  \end{center}\end{minipage}
\begin{minipage}{4.8cm}\begin{center} \includegraphics[width=4.15cm]{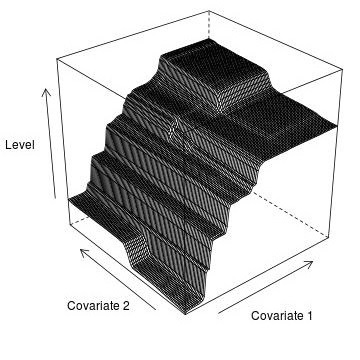}  \end{center}\end{minipage}
           
\caption{True underlying functions (first column) and posterior mean estimates obtained for $\omega=0$ (second column) and $p=-1,q=1$ (third column) for the five regions with neighbourhood structure as detailed in Network 1.}
\label{fig:Sim33a}
\end{center}
\end{figure}

Study 1 considers the geographical network illustrated in Figure \ref{fig:Sim33a1} with region 2 being neighbour of regions 1 and 3, region 3 being neighbour of 2 and 4, etc. The true underlying functions are constructed such that (i) neighbouring functions share similarities and (ii) region 1 and 5 are quite different. Figure \ref{fig:Sim33a} illustrates the true functions (Column 1) and shows, for instance, an increase in the maximum level from region 1 through to 5. The covariate spaces $X_1,\ldots,X_5$ are set equal to the unit square and 300 covariate values are sampled for each region. BSMMR is performed with two settings, (1) $p=1,q=1$ and (2) $p=-1,q=1$, and results are again compared to those for $\omega=0$. 

Table \ref{tab:Sim33} shows an overall improvement in both the MAE and the standard deviation. However, the degree of improvement is not consistent across all five regions. In particular, the model fit improves for regions 2 to 4 while regions 1 and 5 show only small, if any, improvement. The setting $p=-1,q=1$ performs best, based on the average MAE across the five regions, which aligns with the results in Subsection \ref{sec:SimGausspq} since the lower functional levels are more similar than are the upper ones. The larger improvement for regions 2 through 4 is due to their functional shape being close to the average of their neighbours and thus increased smoothing is preferred. In contrast, higher values for $\omega$ imply that region 1 and region 5 are overly smoothed. With respect to the posterior mean plots, Figure \ref{fig:Sim33a} indicates only slight differences for the settings. 

\begin{figure}
\begin{center}
\parbox{2.8cm}{\hfill}
\parbox{4.8cm}{\begin{center}\hspace{0.2cm} \textbf{Truth}\end{center}} 
\parbox{4.8cm}{\begin{center}\hspace{0.2cm} \textbf{Intersection}\end{center}} 
\parbox{4.8cm}{\begin{center}\hspace{0.2cm} \textbf{Union}\end{center}}\\      
\vspace{-0.1cm}
\parbox{2.8cm}{\textbf{Region 1} }  
\begin{minipage}{4.8cm}\begin{center} \includegraphics[width=4.15cm]{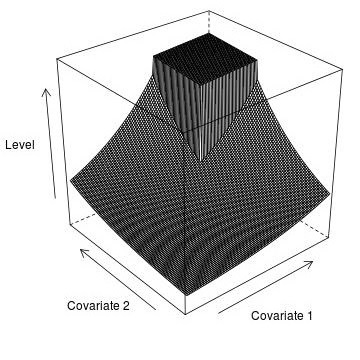} \end{center}\end{minipage} 
\begin{minipage}{4.8cm}\begin{center} \includegraphics[width=4.15cm]{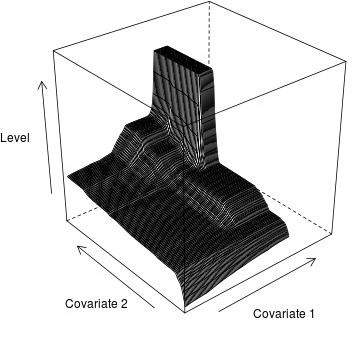} \end{center}\end{minipage}
\begin{minipage}{4.8cm}\begin{center} \includegraphics[width=4.15cm]{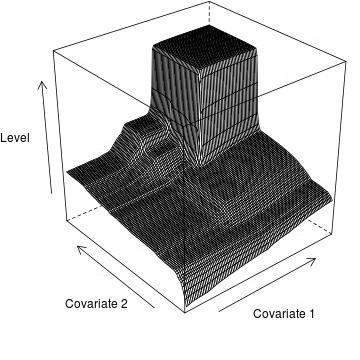}\end{center}\end{minipage}\\
  
\parbox{2.8cm}{\textbf{Region 2} } 
\begin{minipage}{4.8cm}\begin{center} \includegraphics[width=4.15cm]{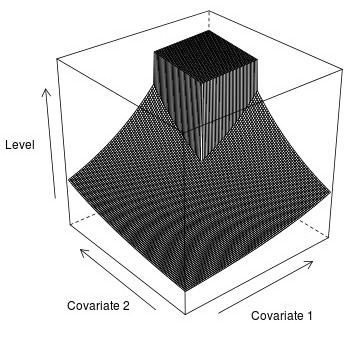} \end{center}\end{minipage} 
\begin{minipage}{4.8cm}\begin{center} \includegraphics[width=4.15cm]{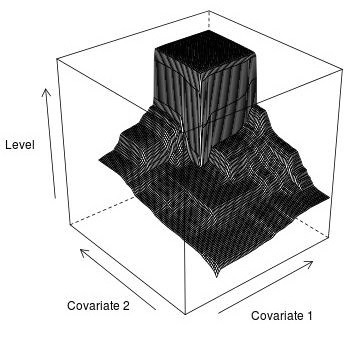} \end{center}\end{minipage}
\begin{minipage}{4.8cm}\begin{center} \includegraphics[width=4.15cm]{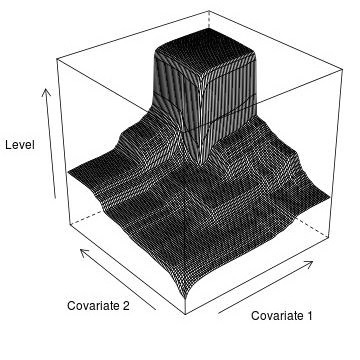}\end{center}\end{minipage}\\
  
\parbox{2.8cm}{\textbf{Region 3} } 
\begin{minipage}{4.8cm}\begin{center} \includegraphics[width=4.15cm]{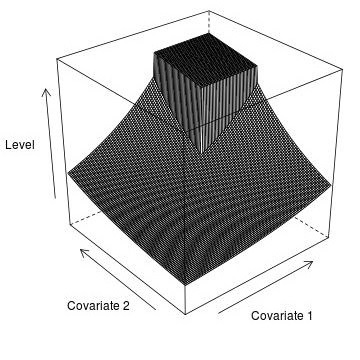}\end{center}\end{minipage}  
\begin{minipage}{4.8cm}\begin{center} \includegraphics[width=4.15cm]{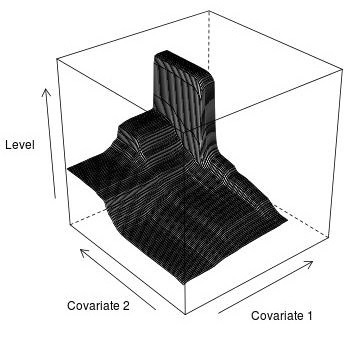} \end{center}\end{minipage}
\begin{minipage}{4.8cm}\begin{center} \includegraphics[width=4.15cm]{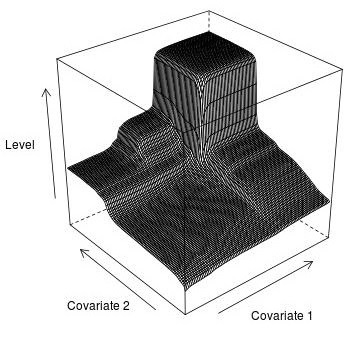}\end{center}\end{minipage}\\
    
\parbox{2.8cm}{\textbf{Region 4} } 
\begin{minipage}{4.8cm}\begin{center} \includegraphics[width=4.15cm]{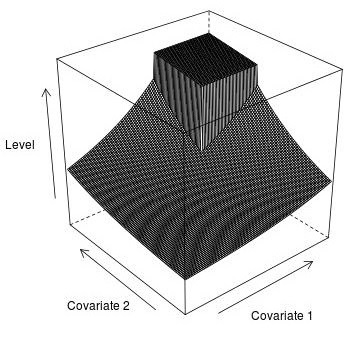}\end{center}\end{minipage} 
\begin{minipage}{4.8cm}\begin{center} \includegraphics[width=4.15cm]{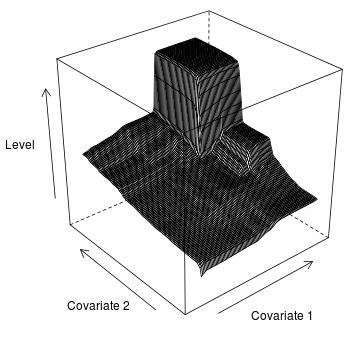} \end{center}\end{minipage}
\begin{minipage}{4.8cm}\begin{center} \includegraphics[width=4.15cm]{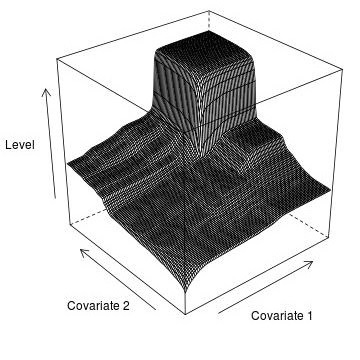}\end{center}\end{minipage}\\
    
\parbox{2.8cm}{\textbf{Region 5} } 
\begin{minipage}{4.8cm}\begin{center} \includegraphics[width=4.15cm]{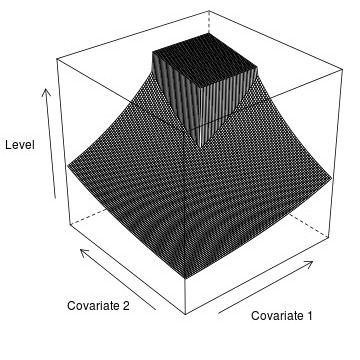}\end{center}\end{minipage} 
\begin{minipage}{4.8cm}\begin{center} \includegraphics[width=4.15cm]{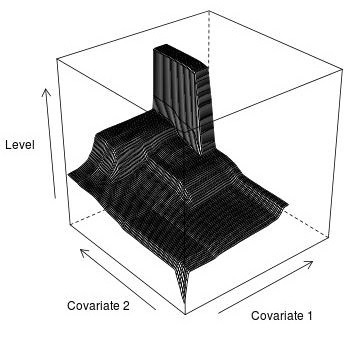}  \end{center}\end{minipage}
\begin{minipage}{4.8cm}\begin{center} \includegraphics[width=4.15cm]{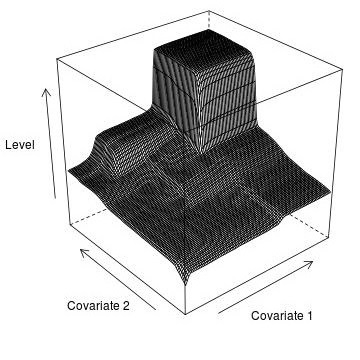}  \end{center}\end{minipage}

\caption{True underlying functions (first column) and posterior mean estimates for the settings in Study 2. The second corresponds to $A_{k,k'}$ being the intersection of $X_k$ and $X_{k'}$ while the last column gives the extrapolated functions for $A_{k,k'}$ being defined as the union of $X_k$ and $X_{k'}$.}
\label{fig:Sim33b}
\end{center}
\end{figure}

The geographical structure for Network 2 is illustrated in Figure \ref{fig:Sim33a2} with regions having between 1 and 4 neighbours. Covariate spaces vary respective to the first covariate, in particular, taking values between 0.0 and 0.7 for regions 1, 3 and 5, 0.1 and 0.9 for region 4 and 0.2 to 1.0 for region 2. This setting facilitates examination of BSMMR with respect to extrapolation to both lower and upper functional levels. The number of observations generated for regions 1 through 5 are 100, 500, 200, 300 and 200, respectively. Figure \ref{fig:Sim33b} shows that the true underlying functions, plotted on the unit square, are similar in their functional levels over the whole covariate space. In this Study 2, BSMMR is applied with $p=1,q=1$ and two different settings for the sets $A_{k,k'}$ in \eqref{eq:D}.

\begin{table}
\begin{center}
\caption{Mean absolute errors$\times10^{-2}$ and standard deviation$\times10^{-2}$ of the residuals obtained by BSMMR for networks of five regions with different settings in the prior density. The last column gives the average mean absolute error of the five regions. The last row in Study 2 provides the mean absolute error on the extrapolated space.}
\begin{tabular}{clcccccc}
\hline
Study & Setting    & Region 1    & Region 2   & Region 3    & Region 4   & Region 5   & MAE\Bstrut\\
\hline
1     & $\omega=0$ & 9.0 (11.6)  & 8.1 (11.4) & 8.4 (11.1)  & 9.4 (11.1) & 9.2 (11.3)  & 8.8 \Tstrut\\  
      & $p=1$      & 8.6 (11.0)  & 7.3 (9.9)  & 8.2 (10.8)  & 8.4 (9.8)  & 9.4 (11.4)  & \hspace{-0.7ex}8.4 \\  
      & $p=-1$     & 8.7 (11.0)  & 7.3 (9.8)  & 7.4 (10.1)  & 8.2 (9.7)  & 9.3 (11.6)  & 8.2 \Bstrut\\  

2     & $\omega=0$ & 11.5 (14.4) & 7.4 (11.9) & 9.7 (12.2)  & 8.2 (13.8) & 9.2 (12.1)  & 9.2 \Tstrut\\  
      & Overlap    & 11.4 (13.9) & 7.7 (11.2) & 7.6 (10.2)  & 7.6 (10.8) & 8.1 (11.5)  & \hspace{-0.7ex}8.5 \\  
      & Union      & 12.5 (14.2) & 8.1 (10.4) & 7.5 (10.0)  & 7.4 (10.6) & 7.5 (10.7)  & \hspace{-0.7ex}8.6 \\    
      &            & 12.8 (17.3) & 9.5 (10.8) & 10.1 (13.8) & 8.5 (12.2) & 12.0 (17.1) & 10.6\Bstrut\\    
\hline
\end{tabular}
\label{tab:Sim33}
\end{center}
\end{table}

In the first setting, $A_{k,k'}$ is set equal to the intersection of $X_k$ and $X_{k'}$, $A_{k,k'} = X_k\cap X_{k'}$. Table \ref{tab:Sim33} indicates a reduction in the overall MAE, compared to $\omega=0$, with larger improvements being achieved for region 3 to 5. In conclusion, region 3 through 5 borrow statistical strength from region 2 as a consequence of the higher number of observations for region 2. Further, estimates for $\lambda_2$ are more similar to $\lambda_3$ to $\lambda_5$ than to $\lambda_1$ due to the weights $d_{k,k'}$, implying that only small amounts of statistical information can be borrowed when estimating $\lambda_1$. Additionally, both MAE and standard deviation are quite high for region 1 due to the paucity of data points and the high variance. Figure \ref{fig:Sim33b} also indicates limitations in fitting the functional shapes around the threshold effect due to the small number of observations.

The second setting defines $A_{k,k'}$ as the union of the covariates spaces $X_k$ and $X_{k'}$, $A_{k,k'} = X_k\cup X_{k'}$. This setting allows to extrapolate the functions $\lambda_1, \ldots, \lambda_5$ by borrowing information from adjacent regions. For the original covariate spaces, Table \ref{tab:Sim33} indicates that the setting $A_{k,k'} = X_k\cup X_{k'}$ performs similarly to $A_{k,k'} = X_k\cap X_{k'}$ in terms of the overall MAE but not with respect to the single regions. In particular, the setting $A_{k,k'} = X_k\cap X_{k'}$ provides the lowest MAE for regions 1 and 2 while $A_{k,k'} = X_k\cup X_{k'}$ reduces the MAE for region 5. Since statistical information for extrapolating $\lambda_2$ and $\lambda_4$ to the lower values of Covariate 1 is borrowed from regions 1, 3 and 5, the model fit on the extrapolated spaces depends on the degree of similarity. Similar arguments hold for the extrapolation of $\lambda_1$, $\lambda_3$ and $\lambda_5$ to the higher values of the first covariate. This effect is observable in the last column of Figure \ref{fig:Sim33b}. The higher values for the MAE in Table \ref{tab:Sim33} on the extrapolated spaces are hence expected.

In summary, all simulations performed in this section demonstrated that BSMMR leads to improved model fit, irrespective of the functional shapes, if similarities between neighbouring regions exists. This conclusion is found for both continuous and discrete observations and independent of the variance of the data process. The proposed \textit{EGO-CV} algorithm, Algorithm \ref{alg:BSMMR}, proved to be suitable and robust for both similar and dissimilar neighbouring functions. While simulations showed a clear sensitivity with respect to $p$ and $q$, the results show little, or no, sensitivity to the model complexity parameter $\eta$. Further, sensitivity with respect to $p$ and $q$ also depends on the distribution of the covariate observations over the associated covariate spaces $X_1,\ldots,X_K$. Finally, Study 2 in Section 3.2 clearly demonstrated the potential of BSMMR to extrapolate monotonic functions.
\section{Case Study}
\label{sec:casestudy}

\begin{figure}
\begin{center}
\includegraphics{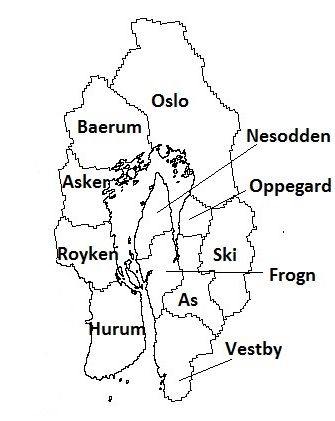}
\caption{Map of the 11 municipalities considered.}
\label{fig:neighbourhood}
\end{center}
\end{figure}

BSMMR is applied in order to investigate the weather dynamics leading to property insurance claims in Norway. The data provide the daily number of claims per Norwegian municipality from 1997 to 2006 due to precipitation, surface water, snow melt, undermined drainage, sewage back-flow or blocked pipes. The monthly number of policies and daily observations of multiple weather covariates are also available for each municipality. While \citet{Scheel2013} consider several covariates, we focus analysis on the amount of precipitation on the current and previous days. This selection is based on \citet{Scheel2013} finding that these are the most important covariates. Intuitively, the claim risk per property increases with the amount of precipitation. Therefore, the monotonicity assumption appears reasonable and BSMMR is applicable; the assumption could be verified based on the test by \citet{Bowman1998}. Analysis is performed for a contiguous set of $K=11$ municipalities around the Oslofjord and Figure \ref{fig:neighbourhood} illustrates the neighbourhood structure.

The applied modelling framework is formalised in the following. Let $N_{k,t}$ denote the number of claims recorded on day $t$ for municipality $k$. Further, $R_{k,t}$ and $R_{k,t-1}$ refer to the amount of precipitation on the current and previous day respectively for municipality $k$. Interest lies in the association between $N_{k,t}$ and the two covariates $R_{k,t}$ and $R_{k,t-1}$. Due to the number of policies $A_{k,t}$ being known, $N_{k,t}$ is modelled by a Binomial distribution with the claim probability $p_{k,t}$ changing monotonically in $R_{k,t}$ and $R_{k,t-1}$. Formally, the claim model is given by 
\begin{equation}
\begin{split}
N_{k,t} &\sim \mbox{Binomial}\left(A_{k,t}, p_{k,t}\right)\\
\mbox{logit}\left(p_{k,t}\right) &= \alpha_k + \lambda_k\left(R_{k,t}, R_{k,t+1}\right),
\end{split}
\end{equation}
where $\lambda_k$ and $\alpha_k$, $k=1,\ldots,11$, are, respectively, the unknown regional monotonic functions and baseline levels. As in the simulation study in Subsection \ref{sec:SimGausseta}, a CAR prior is set on the regional intercepts $\alpha_1, \ldots, \alpha_{11}$.

BSMMR is then applied with prior parameters $p=-1$, $q=1$, $\eta=2$ and $d_{k,k'}=1$ if municipalities $k$ and $k'$ share a border and 0, otherwise. The selection $p=-1$ is due to 
the high occurrence of days with little or no precipitation. More specifically, relatively more data points are available to model the lower functional levels compared to the number of days with high amount of precipitation. Hence, $p=-1$ is a preferable choice based on the simulation findings in Subsection \ref{sec:SimGausspq}. The set $A_{k,k'}$ in the prior density is defined as the union of the covariate spaces for municipality $k$ and $k'$. In order to obtain more uniformly distributed covariate values, BSMMR is applied to the transformed covariates $\sqrt{R_{k,t}}$ and $\sqrt{R_{k,t-1}}$.

Observations for 2001 and 2003 are stored as test data in order to assess and compare predictive performance. The model parameters are then estimated based on the remaining 8 years. BSMMR is compared to two competing models:
\begin{enumerate}
\item Average dailly number of claims over the training data set
\item Geographically varying coefficient (GVC) model with unique CAR priors on each covariate effect.
\end{enumerate}

The \textit{EGO-CV} algorithm, Algorithm 1, is applied with 3 repetitions of 10-fold cross-validation. The initial baseline level is set to -9.0 for all municipalities, based on the GVC model fit for Oslo. The boundaries on the functional levels, $\delta_{\min}$ and $\delta_{\max}$, are set to 0.0 and 6.0, respectively. After selecting an appropriate smoothing parameter $\omega_{opt}$, the final RJMCMC algorithm runs for 1,000,000 iteration steps and every 500th sample is stored for analysis after a burn-in of 200,000. The GVC model is fitted with the two covariate effects by performing 10,000 iteration steps with a burn-in of 1,000. 

\begin{table}
\begin{center}
\caption{Sum of squared prediction errors for 2001 and 2003 based on the model fitted with explanatory variables $R_t$, $R_{t+1}$ for the remaining years between 1997 and 2006.}
\label{tab:CaseStudy}
\begin{tabular}{|l|c|c|c|c|}
\hline
Municipality & Average            & \hspace{1cm}GVC\hspace{1cm}           & BSMMR: $\omega=0$ & BSMMR: $\omega=\omega_{opt}$ \\
\hline
Nesodden     & 20.52              & $\mathbf{19.93}$                      & 20.93             & 20.43              \\
Frogn        & 8.46               & 8.60                                  & $\mathbf{8.31}$   & 8.44               \\
Oppegard     & $\mathbf{26.16}$   & 28.29                                 & 31.92             & 29.64              \\
As           & $\mathbf{13.91}$   & 14.09                                 & 13.95             & 14.07              \\
Vestby       & $\mathbf{18.51}$   & 18.70                                 & 18.61             & 18.61              \\
Ski          & $\mathbf{38.22}$   & 38.70                                 & 39.03             & 39.04              \\
Asker        & 372.55             & $\mathbf{317.93}$                     & 360.72            & 360.80             \\
Baerum       & 915.10             & 663.25                                & 507.80            & $\mathbf{478.82}$  \\
Oslo         & $\mathbf{412.17}$  & 440.27                                & 452.08            & 443.70             \\
Royken       & 63.51              & $\mathbf{53.11}$                      & 58.46             & 58.22              \\
Hurum        & 17.68              & 17.62                                 & 17.53             & $\mathbf{17.49}$   \\
\hline
$\Sigma$     & 1906.8             & 1620.5                                & 1529.3            & $\mathbf{1489.3}$  \\
\hline
\end{tabular}
\end{center}
\end{table}

Table \ref{tab:CaseStudy} shows that BSMMR performs the best in terms of the overall predictive error, denoted $\Sigma$, reducing the value from 1620 for the GVC fit to 1490. In case of Baerum, BSMMR leads to a much smaller predictive error than the GVC fit which indicates that the underlying regression function is non-linear. The difference for Asker is due to one day with a high number of claims that is not captured well as days with similar covariate values in the training data show no occurrence of this magnitude. No substantial differences are found for several municipalities which corresponds to zero high-claim days being observed over the test period. The occurrence of zero high-claim days can be seen in the results as the predictive squared error obtained for the average daily number of claims is low for most municipalities. Finally, the results show that estimates improve by accounting for geographical dependency. The small level of improvement from $\omega=0$ to $\omega=\omega_{opt}$ can be explained by the high number of training data points ($\approx$ 3000) for each municipality. Hence, important structures in the regression surface are likely to be detected without using statistical information from neighbouring municipalities. In conclusion, the application of BSMMR improved the model fit.
\section{Discussion}

We have developed new non-parametric Bayesian methodology which facilitates the modelling and estimation of geographically varying monotonic regression functions. Each regional function is defined to be of piecewise constant form and is represented by a set of marked point processes. Statistical information is 'shared' geographically by a prior which also includes a penalty for model complexity. The prior is constructed based on a pair-wise discrepancy measure which penalizes differences in the functional levels. The discrepancy measure is flexible and allows the geographical dependency to vary with the functional levels. As the normalising constant of the prior was intractable, we developed the \textit{EGO-CV} algorithm, which combines cross-validation and Bayesian global optimization, in order to optimise the smoothing parameter $\omega$. Our simulation and case studies have illustrated that BSMMR has the potential to improve estimates if similarities between neighbouring functions exist. These conclusions were irrespective of the functional shapes and the distribution of the covariate values.

From a general perspective, BSMMR provides a useful modelling approach which allows for both smooth and discontinuous functional forms which may not be captured if a linear or additive form is assumed. The approach may be applied generally for network and dependency modelling and is not limited to a geographical context. BSMMR is well suited for covariate spaces of low dimensions and offers great flexibility. More caution is, however, recommended for higher dimensions, as it is for many flexible modelling approaches. This is due to the computational cost for calculating the integral in the prior distribution scales exponentially with the dimension of the covariate space. Additionally, the monotonic constraint becomes less restrictive with increasing dimensions, leading to a potential overfit of the data. Hence, the considered monotonic functions should preferably be defined on covariate spaces of dimension two to five. Nevertheless, it is possible to estimate an additive model of one monotonic and one linear function or to combine several monotonic functions. This may be in particular useful if two covariate subsets affect the probability model independently from each other. Consequently, the methodology introduced in this paper may be used for higher dimensional covariate spaces but requires a pre-analysis in order to achieve optimal results. 

Computationally, the approach is demanding, depending on the geographical network structure, the dimension of the covariate space and the number of data points per region. As mentioned in the previous paragraph, the calculation of the prior ratio is computationally expensive, as the evaluation of the integral is non-trivial. We reduce the computational time by firstly deriving the area of the covariate space affected by the proposal and then evaluating $D_{p,q}$ over this subspace. However, further splits into smaller subspaces are usually required as the neighbouring functions are likely to vary over this subspace. Another computational step is the update of the likelihood function as monotonicity has to be checked respectively for each data point. Hence, each $s$-fold cross-validation requires a long time. The combination of cross-validation and Bayesian optimization reduced the computational time as the \textit{EGO} algorithm is fast. \texttt{C++} code and \texttt{R} files used in Section \ref{sec:Sim} can be downloaded from \texttt{www.lancaster.ac.uk/pg/rohrbeck/BSMMR}. 

The work presented in this paper can be extended in several ways. From a theoretical perspective, an approach for estimating suitable values for $p$ and $q$ is of interest and will be considered in future research. Alternatively, the discrepancy measure in \eqref{eq:D} may be defined differently, e.g.\ based on the Kullback-Leiber divergence. Further, the decision to perform 10-fold cross-validation in all studies was an arbitrary selection. As the value for $\omega$ depends also on the number of data points, it may be better to have more folds in order to obtain more accurate estimates. The additional computational time may be tackled using parallelised computing techniques. We plan to implement this in future versions of our software package. Finally, the claim model in Section \ref{sec:casestudy} is not the most sophisticated as \citet{Scheel2013} find, for instance, that drainage also effects the claim probability positively. The model also takes no temporal variation into account. The modelling framework is, however, extendible to a spatio-temporal setting with a function being estimated for each municipality for each year. 

\newpage 
 
\subsection*{Acknowledgements}

Rohrbeck gratefully acknowledges funding of the EPSRC funded STOR-i centre for doctoral training (grant number EP/H023151/1). This research was also financially supported by the Norwegian Research Council. This paper greatly benefited from discussions with Jonathan Tawn, Elija Arjas, Christopher Nemeth, Sylvia Richardson, Lawrence Bardwell, Jamie Fairbrother, David Hofmeyr and Ida Scheel. We also thank Ida Scheel for providing access to the Norwegian insurance and weather data.  
\appendix

\section{Details of the RJMCMC algorithm}
\label{sec:AppA}

The RJMCMC algorithm in Section \ref{sec:Inference} outlines our approach to sample realizations from the posterior 
\begin{equation*}
\pi\left(\lambda_1, \ldots, \lambda_K, \bm{\theta}_1, \ldots, \bm{\theta}_K|\mathcal{D}, \omega, \eta\right) \propto \prod_{k=1}^K \prod_{t=1}^{T_k} f\left(y_{k,t}| \lambda_k\left(\mathbf{x}_{k,t}\right), \bm{\theta}_k \right) \times \pi\left(\lambda_1, \ldots, \lambda_K|\omega, \eta\right) \times \pi(\bm{\theta}_1, \ldots, \bm{\theta}_K),
\end{equation*}
where $\mathcal{D}$ denotes the observations of the explanatory and response variables, $\mathbf{x}_{k,t}$ and $y_{k,t}$ respectively, for the $K$ regions. The number $T_k$ refers to the number of observations for region $k$. In the following, the update of one of the monotonic functions is described in detail. 

Function $\lambda_k$, $k=1,\ldots,K$, is updated by selecting one of the subprocesses $\Delta_{k,i}$, $i=1,\ldots,I$ with equal probability and sampling one of the predefined moves. For simplicity in the notation, the probability of proposing \textit{Birth} and \textit{Death} are set to be equal in the following. Here, \textit{Death} or \textit{Shift} are rejected instantly if the process $\Delta_{k,i}$ is empty and, similarly, \textit{Birth} results in an unchanged process if the maximum number of points, $n_{max}$, is reached. The Jacobian in the acceptance probability is equal to 1 as the mapping for adding a point is equal to the identity function. Hence, the acceptance probability for accepting the new function $\lambda_k^*$ conditional on all other functions, $\lambda_{-k}$, is given by 
\begin{equation*}
\begin{split}
\alpha\left(\lambda_k,\lambda_k^{*}\right) &= \min\left\{1,~R\left(\lambda_k,\lambda_k^{*}\right)\right\} \\
&= \min\left\{1,~\prod_{t=1}^{T_k}\frac{f\left(\mathbf{y}_{k,t}~|~\lambda_k^{*}(\mathbf{x}_{k,t}), \bm{\theta}_k\right)}{f\left(\mathbf{y}_{k,t}~|~\lambda_k(\mathbf{x}_{k,t}), \bm{\theta}_k\right)} \times \frac{\pi\left(\lambda_{k}^{*}~|~\lambda_{-k},\omega,\eta \right)}{\pi\left(\lambda_{k}~|~ \lambda_{-k},\omega,\eta \right)}\times \frac{q\left(\lambda_{k}|\lambda_{k}^{*}\right)}{q\left(\lambda_{k}^{*}|\lambda_{k}\right)}\right\},\\
\end{split}
\end{equation*}
where $\mathbf{y}_k=\left(y_{k,1}, \ldots,y_{k,T_k}\right)$ and $\mathbf{x}_k=\left(\mathbf{x}_{k,1}, \ldots,\mathbf{x}_{k,T_k}\right)$ refer to the observed responses and covariate values for region $k$ respectively. 

In case of a \textit{Birth}, the addition of a support point $\left(\bm\xi^*, \delta^*\right)$ is proposed with the location $\bm{\xi}^*$ being sampled uniformly from the space $X_{k,i}$ associated to $\Delta_{k,i}$. The level $\delta^*$ is then sampled uniformly on the interval $\left[b_l,b_u\right]\subseteq\left[\delta_{\min}, \delta_{\max}\right]$ of possible levels such that monotonicity is preserved. The reverse move \textit{Death} selects one of the $n\left(\Delta_{k,i}\right)$ existing points with equal probability and proposes to remove it. Hence, $R\left(\lambda_k,\lambda_k^{*}\right)$ in the acceptance probability result in 
\[R\left(\lambda_k,\lambda_k^{*}\right) = \prod_{t=1}^{T_k}\frac{f\left(\mathbf{y}_{k,t}~|~ \lambda_k^{*}(\mathbf{x}_k), \bm{\theta}_k\right)}{f\left(\mathbf{y}_{k,t}~ |~\lambda_k(\mathbf{x}_k), \bm{\theta}_k\right)} \times \prod_{\substack{k'=1\\k'\neq k}}^K \frac{\exp\left[-\omega \cdot d_{k,k'} \cdot D_{p,q}\left(\lambda_k^{*}, \lambda_{k'}\right)\right]} {\exp\left[-\omega \cdot d_{k,k'} \cdot D_{p,q}\left(\lambda_k, \lambda_{k'}\right)\right]} 
\left(1-\frac{1}{\eta}\right)\times \frac{\left|X_{k,i}\right|\left(b_u-b_l\right)}{n\left(\Delta_{k,i}\right)+1},\]
for a new support point $\left(\bm{\xi}^*, \delta^*\right)$, where $\left|X_{k,i}\right|$ denotes the volume of $X_{k,i}$. Equivalently, it yields to
\[R\left(\lambda_k,\lambda_k^{*}\right) = \prod_{t=1}^{T_k}\frac{f\left(\mathbf{y}_{k,t}~|~ \lambda_k^{*}(\mathbf{x}_k), \bm{\theta}_k\right)}{f\left(\mathbf{y}_{k,t}~|~\lambda_k(\mathbf{x}_k), \bm{\theta}_k\right)}
\times \prod_{\substack{k'=1\\k'\neq k}}^K \frac{\exp\left[-\omega \cdot d_{k,k'} \cdot D_{p,q}\left(\lambda_k^{*}, \lambda_{k'}\right)\right]} {\exp\left[-\omega \cdot d_{k,k'} \cdot D_{p,q}\left(\lambda_k, \lambda_{k'}\right)\right]} \frac{1}{\left(1-\frac{1}{\eta}\right)}\times \frac{n\left(\Delta_{k,i}\right)}{|X_{k,i}|\left(b_u-b_l\right)}\]
for removing a current support point.

Finally, a \textit{Shift} proposes a local change of $\lambda_k$ by shifting both the location and level of an existing support point  but without changing the current partial ordering of the support points. In this work, the index $j^*$ of the support point  $\left(\bm{\xi}_{k,i,j^*}, \delta_{k,i,j^*}\right)$ to be moved is selected with probability proportional to the current level in order to improve the convergence for the higher levels. The new location $\bm{\xi}^*_{k,i,j^*}$ is then sampled uniformly with the lower and upper bounds in each covariate being given by the next higher and lower covariate values; see Appendix of \citet{Saarela2011} for details. The proposed level $\delta_{k,i,j^*}^*$ is then sampled uniformly on the set of possible values which preserve the monotonic constraint. Therefore,
\[R\left(\lambda_k,\lambda_k^{*}\right) = \prod_{t=1}^{T_k}\frac{f\left(\mathbf{y}_{k,t}~|~ \lambda_k^{*}(\mathbf{x}_k), \bm{\theta}_k\right)}{f\left(\mathbf{y}_{k,t}~|~\lambda_k(\mathbf{x}_k), \bm{\theta}_k\right)}
\times \prod_{\substack{k'=1\\k'\neq k}}^K \frac{\exp\left[-\omega \cdot d_{k,k'} \cdot D_{p,q}\left(\lambda_k^{*}, \lambda_{k'}\right)\right]} {\exp\left[-\omega \cdot d_{k,k'} \cdot D_{p,q}\left(\lambda_k, \lambda_{k'}\right)\right]} \times\frac{\delta^*_{k,i,j^*}\sum_{j=1}^{n(\Delta_{k,i})} \delta_{k,i,j}} {\delta_{k,i,j^*}\sum_{j=1}^{n(\Delta_{k,i})} \delta_{k,i,j}^*},\]
where $\sum_{j=1}^{n(\Delta_{k,i})}\delta_{k,i,j}$ denotes the sum of levels of the current support points in process $\Delta_{k,i}$ and $\sum_{j=1}^{n(\Delta_{k,i})} \delta_{k,i,j}^*$ is the updated sum given the proposal.

\section{Details of the posterior mean plots for Study 1 and 4 in Subsection \ref{sec:SimGausseta}}
\label{sec:AppB}

\begin{figure}
\begin{center}
\parbox{2.2cm}{\hfill}
\parbox{5cm}{\begin{center} Truth - Region 2 \end{center}}
\parbox{5cm}{\begin{center} $\omega=\omega_{opt}$ \end{center}}
\parbox{5cm}{\begin{center} $\omega=0$ \end{center}}\\

\parbox{2.2cm}{Study 1}
\begin{minipage}{5cm} \begin{center} \includegraphics[width=5cm]{Simulations/Summary/Sim11/eta=10/omega=0/Analysis/Region2_true_function.jpeg} \end{center}\end{minipage} 
\begin{minipage}{5cm} \begin{center} \includegraphics[width=5cm]{Simulations/Summary/Sim11/eta=10/BayesOptim/Analysis/Region2_function.jpeg} \end{center}\end{minipage} 
\begin{minipage}{5cm} \begin{center} \includegraphics[width=5cm]{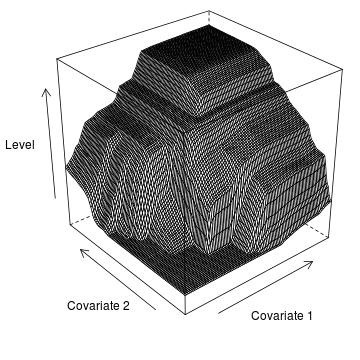} \end{center}\end{minipage}\\

\parbox{2.2cm}{Study 4}
\begin{minipage}{5cm} \begin{center} \includegraphics[width=5cm]{Simulations/Summary/Sim14/eta=2/omega=0/Analysis/Region2_true_function.jpeg} \end{center}\end{minipage} 
\begin{minipage}{5cm} \begin{center} \includegraphics[width=5cm]{Simulations/Summary/Sim14/eta=10/BayesOptim/Analysis/Region2_function.jpeg} \end{center}\end{minipage} 
\begin{minipage}{5cm} \begin{center} \includegraphics[width=5cm]{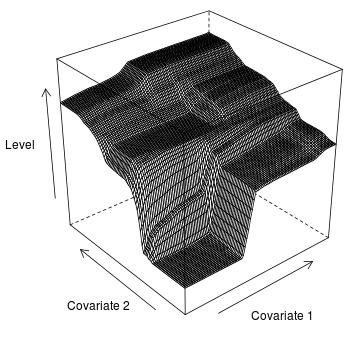} \end{center}\end{minipage}
\caption{True functions (first column) and posterior mean for Region 2 in Study 1 (top row) and 4 (bottom row) obtained by BSMMR with $\eta=10$ and, $\omega=\omega_{opt}$ (second column) and $\omega=0$ (third column), respectively.}
\label{fig:Sim31a}
\end{center}
\end{figure}

The improvements in Table \ref{tab:Sim31} for Region 2 can also be visualised for Study 1 and 4. Study 1 considered the case of both monotonic functions being identical and continuous. Figure \ref{fig:Sim31a} illustrates that the posterior mean for $\lambda_2$ appears smoother for $\omega=\omega_{opt}$ than for $\omega=0$, in particular, for the higher functional levels of $\lambda_2$. For Study 4 with $\lambda_1$ and $\lambda_2$ being similar, Figure \ref{fig:Sim31a} shows an improved modelling of the threshold effect. The posterior mean for $\omega=0$ shows no clear location of the threshold effect and the surface appears slightly continuous. For $\omega=\omega_{opt}$, the threshold effect is fitted better and its location is much better visible. In summary, the posterior mean plots illustrate that the estimates obtained for $\omega=\omega_{opt}$ may improve both the estimation of smooth surfaces as well as threshold effects.

\bibliography{myrefs}
\end{document}